\documentclass[reprint,fleqn,superscriptaddress,twocolumn,showpacs,preprintnumbers,amsmath,amssymb,prb,aps]{revtex4-1}

\usepackage{amsmath} 
\usepackage{amssymb} 
\usepackage{makeidx} 
\usepackage{graphicx} 
\usepackage{epstopdf}
\usepackage{textcomp}
\usepackage{verbatim}
\usepackage{hyperref} 
\usepackage{subfigure}
\usepackage{color}
\usepackage{nicefrac}

\begin{document}
\title{The $SU(4)-SU(2)$ crossover and spin filter properties of a double quantum dot nanosystem}
\author{V.~Lopes}
\email{victor.lopes@aluno.puc-rio.br}
\affiliation{Departamento de F\'{\i}sica, Pontif\'{\i}cia Universidade Cat\'olica do Rio de Janeiro (PUC-Rio),RJ, 22453-900, Brazil}
\author{R.~A.~Padilla}%
\email{ronald@ele.puc-rio.br}
\affiliation{Departamento de F\'{\i}sica, Pontif\'{\i}cia Universidade Cat\'olica do Rio de Janeiro (PUC-Rio),RJ, 22453-900, Brazil}
\author{G.~B.~Martins}
\affiliation{Instituto de F\'{\i}sica, Universidade Federal Fluminense, 24210-346 Niter\'oi, RJ, Brazil}
\author{E.~V.~Anda}
\affiliation{Departamento de F\'{\i}sica, Pontif\'{\i}cia Universidade Cat\'olica do Rio de Janeiro (PUC-Rio),RJ, 22453-900, Brazil}
\date{\today}%
\begin{abstract}
The $SU(4)-SU(2)$ crossover, driven by an external magnetic field $h$, is analyzed in a capacitively-coupled double-quantum-dot device connected to independent leads. As one continuously charges the dots from empty to quarter-filled, by varying the gate potential $V_g$, the crossover starts when the magnitude of the spin polarization of the double quantum dot, as measured by $\langle n_{\uparrow}\rangle -\langle n_{\downarrow}\rangle$, becomes finite. 
Although the external magnetic field breaks the $SU(4)$ symmetry of the Hamiltonian, the ground state preserves it in a region of $V_g$, where $\langle n_{\uparrow}\rangle -\langle n_{\downarrow}\rangle =0$. 
Once the spin polarization becomes finite, it initially increases slowly until a sudden change occurs, in which $\langle n_{\downarrow}\rangle$ (polarization direction opposite to the magnetic field) reaches a maximum and then decreases to negligible values abruptly, at which point an orbital $SU(2)$ ground state is fully established. 
This crossover from one Kondo state, with emergent $SU(4)$ symmetry, where spin and orbital degrees of freedom all play a role, to another, with $SU(2)$ symmetry, where only orbital degrees of freedom participate, is triggered by a competition between $g\mu_Bh$, the energy gain by the Zeeman-split polarized state and the Kondo temperature $T_K^{SU(4)}$, the gain provided by the $SU(4)$ unpolarized Kondo-singlet state. 
At fixed magnetic field, the knob that controls the crossover is the gate potential, which changes the quantum dots occupancies. 
If one characterizes the occurrence of the crossover by $V_g^{max}$, the value of $V_g$ where $\langle n_{\downarrow}\rangle$ reaches a maximum, one finds that the function $f$ relating the Zeeman splitting, $B_{max}$, that corresponds to $V_g^{max}$, i.e., $B_{max}=f\left(V_g^{max}\right)$, has a similar universal behavior to that of the function relating the 
	Kondo temperature to $V_g$.
In addition, our numerical results show that near the $SU(4)$ Kondo temperature and for relatively small magnetic fields the device has a ground state that restricts the electronic population at the dots to be spin polarized along the magnetic field. 
These two facts introduce very efficient spin-filter properties to the device, also discussed in detail in the paper. 
This phenomenology is studied adopting two different formalisms: the Mean Field Slave Bosons Approximation, which allows an approximate analysis of the dynamical properties of the system, and a Projection Operator Approach, which has been shown to describe very accurately the physics associated to the ground state of Kondo systems.
\end{abstract}
\maketitle

\section{Introduction}\label{sec1}

The discovery in 1998 of the Kondo effect in artificial atoms \cite{Kastner1998}, so-called quantum dots (QDs), has greatly motivated the study of this phenomenon in nanostructures in the last two decades. Since the pioneering works where QDs were shown to possess all the properties of real atoms \cite{Kouwenhoven2001}, very many investigations were done to determine the behavior of different structures of QDs associated to the Kondo effect \cite{Pustilnik2004,Choi2006,Al-Hassanieh2009,Chang2009,Florens2011,Gavin2010}. 
It has been shown that nano-systems with QDs are powerful tools to experimentally investigate a variety of properties of highly correlated electrons \cite{Potok2007,Amasha2013,Keller2014,Keller2015}. QDs have proven as well to have very interesting applications as quantum gates \cite{Chiappe2010}, spin filters \cite{Recher2000,Borda2003,Feinberg2004,Hanson2004,Dahlhaus2010,Mireles2006,Hedin2011}, and thermal conductors \cite{Chen2002,Chen2003,Shakouri2006}. 
Transport properties as a function of temperature, magnetic field, and gate potential, have been analyzed in systems with lateral QDs \cite{Goldhaber-Gordon1998,Kogan2004}, carbon nanotubes \cite{Laird2015}, mole\-cular transistors \cite{Natelson2006b}, etc. The major reason the interest in these studies has increased is due to advances in experimental techniques and in the fabrication of nano-devices, which have raised the prospect of many applications in areas like nano-electronics \cite{Goldhaber-Gordon1997,Goodnick2003}, spintronics \cite{Zutic2004}, and quantum computation \cite{Wei2014}.

All this experimental activity has greatly promoted the development of new theoretical studies and formalisms to analyze this phenomenology. A large amount of theore\-ti\-cal predictions related to electronic transport have been obtained using numerical methods. Among the most largely utilized we can mention the Numerical Renormalization Group (NRG)\cite{NRG}, the Density Matrix Renormalization Group (DMRG)\cite{DMRG}, and the Logarithmic Discretized Embedded Cluster Approximation (LDECA)\cite{LDECA}. Other algebraic approaches have as well been used as the various Slave Boson Approximations\cite{SB} and Projection Operator Approach (POA)\cite{POA_1,POA_2} and others based on the Green's function formalism as the Non-Crossing and One-Crossing Approximation (NCA, OCA)\cite{NCA_1,NCA_2,OCA}, and the equation of motion method\cite{Zubarev}. 
In addition, it should be mentioned the use of the Perturbative Renormalized Group Approach \cite{Hewson1993,Hewson2005} as 
well as extensions of Noziere's Fermi liquid-like theories \cite{Mora2008,Mora2015,Filippone2017}. 

In the last years, several studies have appeared in the literature related to the Kondo effect in which, in addition to the spin degree of freedom, the nanostructure presents degenerate orbital degrees of freedom, such that the complete symmetry of the system corresponds to the $SU(N)$ Lie group, for $N>2$. This was the case, for $N=4$, of a single atom transistor \cite{Tettamanzi2012}, in carbon nanotubes \cite{Franceschi2005}, and in capacitively-coupled double QDs \cite{Holleitner2004}. Several theoretical interpretations have been proposed \cite{Lim2006} and, in particular, more closely related to our work, it was theoretically shown that there is an $SU(4)-SU(2)$ crossover when the $SU(4)$ symmetry is broken by either introducing a different gate voltage $V_{g}$ in each dot or by connecting them to the leads by different hopping matrix elements $V$. In addition, it was shown that by manipulating the parameters of the system, without explicitly restoring the broken symmetry, the ground state might 
display, as an `emergent' property, the $SU(4)$ symmetry \cite{Tosi2013} (see also Ref.~\onlinecite{Keller2014}).
Here, it should be noticed that Ref.~\onlinecite{Nishikawa2016} has shown that the conclusions 
of Tosi \emph{et al.}~\cite{Tosi2013} regarding the emergent $SU(4)$ symmetry are 
asymptotically achieved if the intradot and interdot Coulomb repulsions 
are larger than the half-bandwidth (see also Ref.~\onlinecite{Nishikawa2013}). 

Finally, recent theoretical studies of a double quantum dot (DQD) device, connected to two 
independent channels, under the effect of a magnetic field, was shown to exhibit 
an exotic $SU(2)$ Kondo state with the property of having spin polarized currents 
(of opposite polarization) through each QD \cite{Busser2012}. 

In this work, we will concentrate on two main subjects: (i) the Kondo $SU(4)-SU(2)$ crossover, driven by 
an external magnetic field $h$, occurring in a capacitively-coupled DQD device and (ii) the associated spin-filter 
properties of this capacitively-coupled DQD device that emerge in the $SU(2)$ side of the crossover. Although 
some aspects of related problems have already been studied (see Ref.~\onlinecite{Busser2014} for (i) 
and Refs.~\onlinecite{Busser2012,Vernek2014} for (ii), and references therein), there are very important 
properties of this crossover that were not analyzed yet and will be discussed here. 

The main ideas behind the $SU(4)-SU(2)$ crossover can be summarized as follows. 
The crossover is driven by the magnetic field $h$ (causing a Zeeman splitting $B$) that decreases the 
symmetry of the Hamiltonian from $SU(4)$ to $SU(2)$. 
Despite the presence of a finite magnetic field, our results show that 
the symmetry of the ground state changes from $SU(4)$ to $SU(2)$ when the gate potential 
applied to the DQD is reduced.
That the ground state of the DQD may have a higher symmetry, $SU(4)$, than its 
$SU(2)$-symmetric Hamiltonians \textemdash\, a manifestation of an effect dubbed an `emergent' 
$SU(4)$ Kondo ground state \cite{Keller2014} \textemdash\, 
is by itself an interesting result. 

Indeed, we show in Sec.~\ref{sec4}, that the $SU(4)-SU(2)$ crossover can be studied by taking the value of the 
spin polarization, i.e., the difference $\langle n_{\uparrow}\rangle -\langle n_{\downarrow}\rangle$, 
evaluated in the ground state of the DQD system, as playing a similar role to an order parameter that 
defines the transition between two phases, although in this case we are dealing with a crossover process. 
At a particular value of the external field $h$, which produces a Zeeman splitting $B=g\mu_Bh$, 
the crossover is characterized as occurring at the gate potential value 
$V_g^{max}$ where the electronic spin-down occupation, $\langle n_{\downarrow}\rangle$, 
has a very well-defined maximum, denoted $\langle n_{\downarrow}\rangle^{max}$ (see Fig.~\ref{figure4}). 
We name the Zeeman splitting corresponding to this maximum as $B_{max}$. 
If we then analyze the functional relation between $B_{max}$ and $V_g^{max}$, i.e., 
$B_{max}=f\left(V_g^{max}\right)$, our results show that, within the Kondo regime, $f$ has 
a similar universal behavior to that the Kondo temperature has as a function of the gate potential. 
It should be noted that the crossover, as defined here, occurs even when the 
system is deep inside the charge fluctuation regime, in which case it cannot properly be 
said that the system has a Kondo ground state. 
The existence of this clear maximum, irrespective of the regime the system is in, 
allows $B_{max}$ to be characterized as the energy scale controlling the crossover. 

Regarding subject (ii) mentioned above, i.e., the spin-filter properties of the DQD system studied here, 
our results show that, in the $SU(2)$ side of the crossover the 
electronic population at the QDs is already clearly polarized along the magnetic field.
As to the important question, regarding what is the minimum temperature and minimum magnetic field needed 
for the DQD to operate as a spin-filter device, our results show that, as $B_{max}$ is much smaller than the 
Kondo $SU(4)$ temperature ${T_K}^{SU(4)}$, it could operate at temperatures around $10$~K, with a field 
$h \approx 0.1$ Tesla. 
These two facts introduce very efficient spin-filter properties to the device, also discussed in detail in the paper.

\begin{figure}[h!]
\center
\includegraphics[width=\columnwidth]{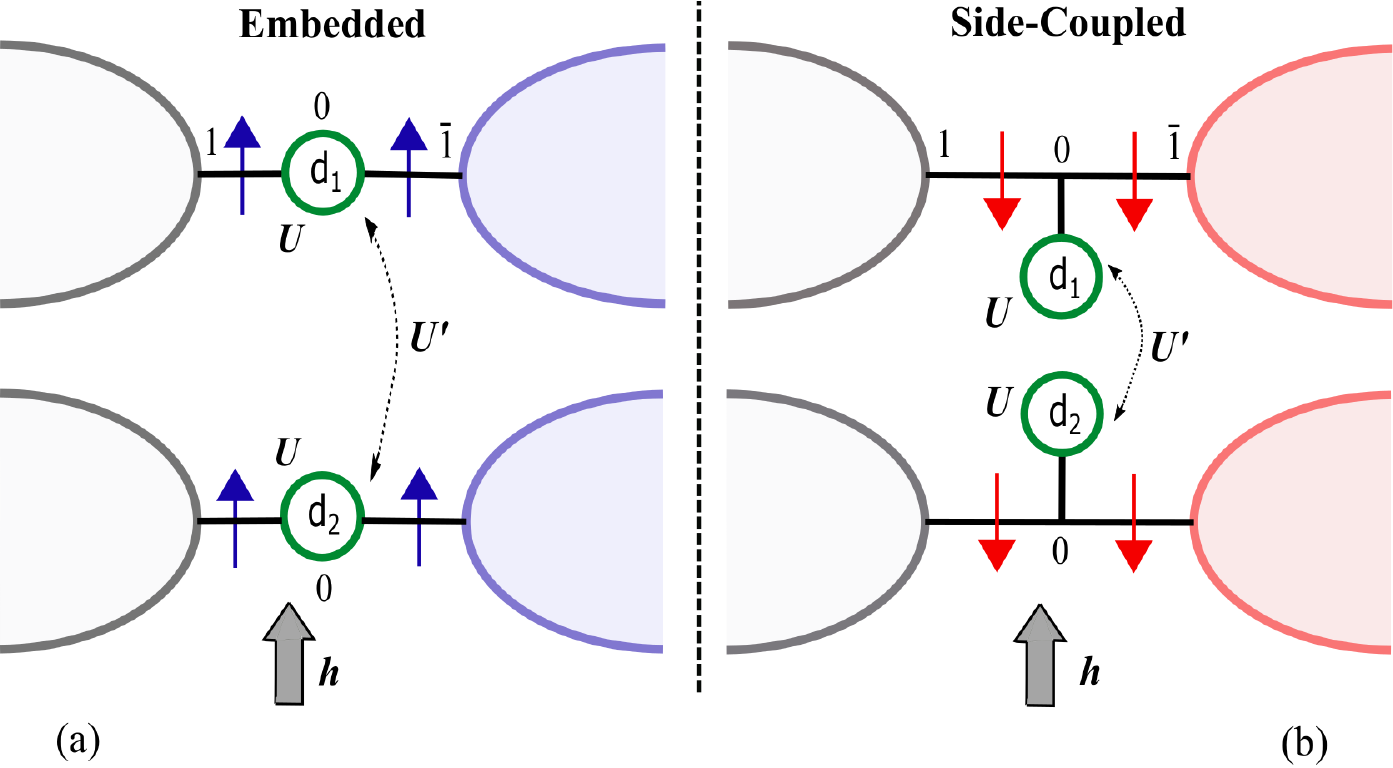}
\caption{(Color online) Capacitively-coupled DQD system. $U^{\prime}$ ($U$) is the inter-QD (intra-QD) Coulomb 
	repulsion, and the system is in the presence of an external magnetic field ${h}$ that acts only in the QDs. 
	In addition, each QD is connected to its adjacent leads by a hopping matrix element $V$ (not shown). 
	The QDs are either embedded (a) into the leads or side-coupled (b) to them. 
	It will be shown in Section V that a relatively small magnetic field can polarize 
	the current transmitted through the QDs with a polarization parallel to the field for embedded QDs 
	and antiparallel to it in the case of side-coupled QDs, as schematically illustrated in 
	panels (a) and (b), respectively. 
	} 
\label{figure1}
\end{figure}

This phenomenology is studied adopting two different formalisms: (i) the Mean Field Slave Bosons Approximation (MFSBA) \cite{sb_1, sb_2, sb_3, sb_4, sb_5, sb_6}, which allows an approximate analy\-sis of the dynamical properties of the system, and (ii) the POA, which has been shown to describe, almost e\-xact\-ly, the static properties associated to the ground state of the Anderson Impurity Hamiltonian \cite{POA_1,POA_2}. Note that we have extended the POA, originally derived to study single-impurity Kondo problems, to the analysis of two capacitively-coupled local levels. As it was the case for single-impurity problems, this extension can be consi\-de\-red to provide almost exact results, as far as the static zero-temperature properties are concerned. In Ref.~\onlinecite{POA_1} the POA results for various Kondo static properties agree quite well with the Bethe Anzats \cite{Bethe_Ansatz} exact results.
It is important to mention that both approaches used to study the system, the MFSBA and the POA, provide the same qualitative and semi-quantitative physical description.

The rest of the paper is organized as follows: In section \ref{sec2}, we provide a description of the capacitively-coupled DQD system; in section \ref{sec3} we present the MFSBA and the POA used to study the properties of the system; section \ref{sec4} is dedicated to the analysis of the $SU(4)-SU(2)$ crossover; section \ref{sec5} describes the spin filter characteristics of the DQD device. We end the paper in section \ref{sec6} with the conclusions. The theoretical methods used are discussed in detail in appendixes A and B.

\section{Description of the system}\label{sec2}

The system is composed by two parallel QDs, each one connected to two independent contacts (see Fig.~\ref{figure1}). 
These QDs, besides an intra-QD Coulomb interaction $U$, are also capacitively coupled by an inter-QD Coulomb interaction $U^{\prime}$. In addition, they are under the influence of an external magnetic field $h$, as shown in Fig \ref{figure1}. 
On one hand, the two configurations shown in panels (a) and (b) in Fig.~\ref{figure1} give identical results from the point of view of the $SU(4)-SU(2)$ crossover and related physics. On the other hand, whether the QDs are embedded [Fig.~\ref{figure1}(a)] or side-coupled [Fig.~\ref{figure1}(b)] to the contacts plays a fundamental role in the transport properties of the system, and the difference 
in these properties will be explicitly analyzed below when we study the conductance. 
The ge\-ne\-ral discussion regarding the $SU(4)-SU(2)$ crossover is presented for the side-coupled QDs geometry \cite{Sasaki2009}. 
Note that similar physics can be obtained using instead a single carbon nanotube QD, where the extra degree of freedom, besides spin, is provided by the valley quantum number present in the graphene honeycomb lattice \cite{Franceschi2005,Laird2015}. 

The system will be described by an extension of the Anderson Impurity Model (AIM) Hamiltonian \cite{Hewson,AIM}, appropriate for two impurities, plus the Zeeman term, given by 
\begin{flalign}
\label{eq:HIM0}
	H_{\rm tot}& = H_{\rm band} + H_{\rm DQD} + H_{\rm hyb} + H_{\rm Zeeman}, &
\end{flalign}
where
\begin{flalign}
\label{eq:HIM1}
	H_{\rm band}&=\sum_{j,\pmb{k}_j,\sigma}\epsilon_{\pmb{k}_j}c_{\pmb{k}_j\sigma}^{\dagger}c_{\pmb{k}_j\sigma}, &
\end{flalign}

\begin{flalign}
\label{eq:HIM2}
	H_{\rm DQD}&=\sum_{j,\sigma}\left( V_{g}n_{j\sigma}+\frac{U}{2}n_{j\sigma}n_{j\bar{\sigma}} \right)+
	U^{\prime}\sum_{\sigma,\sigma^{\prime}}n_{1\sigma}n_{2\sigma^{\prime}}, &
\end{flalign}

\begin{flalign}
\label{eq:HIM3}
	H_{\rm hyb}&=\sum_{j,\pmb{k}_j,\sigma} V_{\pmb{k}_j}\left(c_{dj\sigma}
	c_{\pmb{k}_j\sigma}^{\dagger}+c_{dj\sigma}^{\dagger}c_{\pmb{k}_j\sigma}\right), &
\end{flalign}

\begin{flalign}
\label{eq:HIM4}
	H_{\rm Zeeman}&= \sum_{j} g\mu_BS_j^zh, &
\end{flalign}
where $j=1,2$ labels the QDs and the corresponding attached contacts (after a symmetric/antisymmetric 
transformation between left and right contacts), which are modeled in eq.~\eqref{eq:HIM1} as non-interacting Fermi seas with dispersion $\epsilon_{\pmb{k}_j}$, where $c_{\pmb{k}_j\sigma}^{\dagger}$ $\left( c_{\pmb{k}_j\sigma}\right)$ creates (annihilates) an electron with spin $\sigma$ in contact $j$. Equation \eqref{eq:HIM2} models the QDs, introducing a Coulomb repulsion $U$ between electrons in the same QD, as well as an inter-QD repulsion $U^{\prime}$, where $c_{dj\sigma}^{\dagger}$ $\left( c_{dj\sigma}\right)$ creates (annihilates) an electron with spin $\sigma$ in QD $j$,  $n_{j\sigma}=c_{dj\sigma}^{\dagger} c_{dj\sigma}$ is the number operator in QD $j$, and we assume that the same gate potential $V_{g}$ is applied to each QD. In eq.~\eqref{eq:HIM3}, $V_{\pmb{k}_j}$ couples each QD to the corresponding lead (see Fig.~\ref{figure1}). 
As usual, we take the matrix element $V_{\pmb{k}_j}=V$ to be independent of momentum $\pmb{k}_j$. 
Note that, unless stated otherwise, for the sake of brevity, as $n_{1\sigma}=n_{2\sigma}$, we will from now on drop the $j$ sub-index when referring to the spin occupation number of the QDs. 
Finally, eq.~\eqref{eq:HIM4} describes the effect of an applied magnetic field $h$ acting on spins with magnetic moment $g\mu_B$ in both QDs (where $\mu_B$ is the Bohr magneton and $g$ is the gyromagnetic factor of the electrons in the QD).

Rigorously speaking, at $h=0$, the system only has $SU(4)$ symmetry when both the gate potential $V_g$ and the hybridization matrix element $V$ are independent of $j$, and, in addition, $U^{\prime}=U$. In particular, we assume $U$ and $U^{\prime}$ to be infinite, which restricts the QDs occupations to be either zero or one, a condition that simplifies significantly the numerical calculations. However, within the context of the MFSBA, we consider a case where $U^{\prime}$ is finite in order to show that in the appropriate region of the parameter space the physical properties of the system do not depend upon the particular value of the inter-QD Coulomb repulsion. 

\section{The Mean Field Slave Bosons Approximation and the Projection Operator Approach}\label{sec3}

In this section, we will briefly discuss the two formalisms used to study the properties of the DQD system. A more detailed presentation of these two treatments is given in appendixes A and B. Although most of the discussion is restricted to the case where the inter-QD repulsion is equal to the intra-QD one, i.e., $U=U^{\prime}\rightarrow\infty$, the case of finite $U^{\prime}$ is explicitly treated in the MFSBA calculations. Although, as mentioned above, this could be a more realistic situation, we will see that, in the region of parameter space where $|V_{g}|< U^{\prime}$, the results do not qualitatively depend upon the particular value of $U^{\prime}/U$.

\subsection{Mean Field Slave Bosons Approximation}
As already mentioned we assume that $U\rightarrow\infty$, which simplifies the treatment, as it eliminates double occupied intra-QD states from the Hilbert space. However, as just mentioned above, we will also present results for double inter-QD occupation, taking a finite value for $U^{\prime}$. Following the MFSBA formalism \cite{sb_1, sb_2}, it is necessary to introduce new bosonic operators. As discussed in detail in appendix A, seven auxiliary operators are introduced, each one associated to a different eigenstate of the isolated DQD system, as shown in Table~\ref{table:slavebosons}. 

\begin{table}[ht]
\caption{Eigenstate, eigenenergy, assigned slave-boson (SB) operator, and total number of electrons 
$n_{tot}=\sum_{j,\sigma}n_{j\sigma}$, for the DQD system with Zeeman splitting $B$, 
for $U\rightarrow\infty$ and finite $U^{\prime}$.} 
\centering 
\begin{tabular}{c c c c} 
\hline\hline 
	Eingenstate & Eigenenergy & SB & $n_{tot}$  \\ [0.5ex] 
\hline 
$|0;0\rangle$ & 0 & e & 0 \\ 
$|\uparrow;0\rangle$ & $V_g - B$ & $p^{\uparrow}_{1}$ & 1 \\
$|\downarrow;0\rangle$ & $V_g+ B$ & $p^{\downarrow}_{1}$ & 1 \\
$|0;\uparrow\rangle$ & $V_g - B$ & $p^{\uparrow}_{2}$ & 1 \\
$|0;\downarrow\rangle$ & $V_g + B$ & $p^{\downarrow}_{2}$ & 1 \\
$|\uparrow;\downarrow\rangle$ & $2V_g + U^{\prime}$ & $d^{\uparrow\downarrow}_{12}$ & 2 \\
$|\downarrow;\uparrow\rangle$ & $2V_g + U^{\prime}$ & $d^{\downarrow\uparrow}_{12}$ & 2 \\
$|\uparrow;\uparrow\rangle$ & $2V_g + U^{\prime} - 2B $ & $d^{1}_{\uparrow}$ & 2 \\
$|\downarrow;\downarrow\rangle$ & $2V_g + U^{\prime} + 2B $ & $d^{1}_{\downarrow}$ & 2 \\ [1ex] 
\hline 
\end{tabular}
\label{table:slavebosons} 
\end{table}

A new Hamiltonian can be written with the help of these operators. Restrictions on the Hilbert space are necessary in order to remove additional non-physical states, which is accomplished by imposing relationships among these operators [eqs.~(A1) and (A2)]. The boson operators, within the mean-field approximation \cite{sb_1}, are replaced by their respective expectation values: $e$\textrightarrow $\langle e \rangle$, $p_{\sigma}$ \textrightarrow $\langle p_{j}^{\sigma} \rangle$, $d_{12}$ \textrightarrow $\langle d^{\sigma\bar{\sigma}}_{12} \rangle$, $d_{1\sigma}$ \textrightarrow $\langle d^{1}_{\sigma} \rangle$. The restrictions on the mean values of the bosonic operators are incorporated through the Lagrange multipliers $\lambda$ and $\lambda_{j\sigma}$. 
Following this procedure and in order to simplify the notation,we assume that the bosons operators denote their mean va\-lues. In this case, the effective Hamiltonian can be written: 

\begin{align}
\label{eq:app_H_eff}
H_{eff} &=\sum_{j,k_{j},\sigma}\epsilon_{k_{j}}n_{k_{j},\sigma} +\sum_{j,\sigma}\left( V_{g} - \sigma B\right) c^{\dagger}_{dj,\sigma}c_{dj,\sigma} \nonumber\\
& \quad + U^{\prime}\sum_{\sigma}d^{\sigma\bar{\sigma}\dagger}_{12}d^{\sigma\bar{\sigma}}_{12} + \sum_{\sigma}(U^{\prime}\pm 2B)d^{1\dagger}_{\sigma}d^{1}_{\sigma} \nonumber\\
& \quad + \sum_{{j,\sigma}}V_{j}(c^{\dagger}_{k_{j},\sigma}c_{dj,\sigma} + h.c)Z_{j\sigma} + \lambda(I-1) \nonumber\\
& \quad + \sum_{j,\sigma} \lambda_{j,\sigma}(c^{\dagger}_{dj,\sigma}c_{dj,\sigma} - Q_{j,\sigma}).&
\end{align}

The effective Hamiltonian corresponds to a one-body quasi-fermionic system in which the local energy 
levels in each QD are renormalized by its respective spin dependent Lagrange 
multiplier: $\epsilon_{\sigma} =V_{g} -\sigma B + \lambda _{\sigma}$. As discussed in appendix A, the bosonic operator expectation values and the Lagrange multipliers ($\lambda_{j,\sigma}$\textrightarrow$\lambda_{\sigma}$), necessary to impose the charge conservation conditions, are determined by minimizing the total energy and the free energy of the system. This requires the self-consistent solution of a system of nine equations, thus obtaining the parameters that define the effective one-body Hamiltonian, eq.~\eqref{eq:app_H_eff}, which can then be solved by applying a standard Green's function method.

\subsection{Projection Operator Approach}

The ground state energy, $E$, of our $N$-particle system satisfies the eigenvalue Schr\"odinger equation
\begin{equation}
\label{eq:gstate}
H\vert \Psi\rangle=E\vert \Psi\rangle ,
\end{equation}
where $\vert \Psi\rangle$ represents the ground state eigenvector of the model Hamiltonian, eq.~\eqref{eq:HIM0}. We proceed by projecting its Hilbert space into two subspaces, $S_1$ and $S_2$, and constructing a renormalized Hamiltonian $H_{\rm ren}$ that operates in just one of them \cite{POA_1,POA_2}. For the case of subspace $S_1$, $H_{\rm ren}$ can be written as \cite{Hewson}, 
\begin{equation}
H_{\rm ren}=H_{11}+H_{12}\left(E-H_{22}\right) ^{-1}H_{21},
\end{equation}
where,
\begin{equation}
H_{\rm ij}=\vert i\rangle\langle i\vert H\vert j\rangle\langle j\vert, 
\end{equation}
and state $\vert i \rangle$ belongs to subspace $S_i$. In our case, subspace $S_{1}$ contains only state $\vert 1 \rangle$, consisting of the tensor product of the ground state of the two Fermi seas with the uncharged DQD. All the other states are contained in subspace $S_{2}$, which can be accessed from subspace $S_{1}$ through successive applications of the $H_{21}$ operator. 
It is convenient to define $\Delta E$, as the difference between the ground state energy $E$ and $2\epsilon_{T}$, the sum of the energies of the two uncoupled contact Fermi seas, 

\begin{align}
\label{eq:POA_DeltaE}
\Delta E &=E-2\epsilon_{T},
\end{align}
where $\epsilon_{T}$ is given by 
\begin{align}
\epsilon_{T}&=2\int _{-2t}^{0 }\omega \rho \left(\omega\right) d\omega, 
\end{align}
and $\rho(\omega)$ is the density of states of the Fermi sea.
As shown in appendix B, $\Delta E$ can be found by solving 
\begin{equation}
\label{eq:POA_sol}
\Delta E=f_{1}\left(\Delta E\right),
\end{equation}
where $f_{1}\left(\xi\right)$ and $f_{0}\left(\xi\right)$, given by 
\begin{align}
\label{eq:POA_f1_cont}
f_{1}\left(\xi\right)&=\sum_{\sigma}\int_{-2t}^{0}\bigg \{ \rho \left( \omega \right) \times \nonumber\\
&\dfrac{2V^{2}}{\xi +\omega -V_{g}+\sigma B-f_{0}\left(\xi+\omega\right) }\bigg \}d\omega, 
\end{align}
and 
\begin{equation}
\label{eq:POA_f0_cont}
f_{0}\left(\xi\right)=\int_{0}^{2t}\bigg \{ \rho\left(\omega\right)\dfrac{V^{2}}{\xi-\omega-f_{1}\left(\xi-\omega\right)}\bigg \} d\omega, 
\end{equation}
are obtained self-consistently.  

As briefly described above, the POA results depend on the choice of a convenient $S_1$ subspace, where the model Hamiltonian will be projected, resulting in an effective Hamiltonian. In our case, consisting of two identical QDs with infinite intra-QD Coulomb repulsion, two auxiliary functions have to be self-consistently obtained. Although this requires only a moderate numerical effort, it becomes more involved, and therefore computationally more expensive, in a more general situation of two different QDs and finite intra-QD Coulomb repulsion, as the number of functions to be self-consistently determined increases accordingly.

\section{The $SU(4)-SU(2)$ crossover}\label{sec4}

In this section, we study the $SU(4)-SU(2)$ crossover driven by an external magnetic field applied to the 
DQD system. The QD occupation numbers are used to characterize the crossover. With this objective, 
$\langle n_{\uparrow} \rangle$ and $\langle n_{\downarrow} \rangle$ at each QD is calculated as a 
function of the gate potential using both methods, the MFSBA and the POA. Unless stated otherwise, the parameters taken to 
perform the calculations (in units of $\Delta$, see below) are as follows: 
the coupling between each QD and the corresponding contact is 
$V=8.0$, the half-bandwidth of the contacts is $D=64.0$, and the Zeeman splitting is
given by $B= 3.2 \times 10^{-3}$. Taking ty\-pi\-cal values for GaAs, 
for instance, this corresponds to a magnetic field $h \lesssim 0.1$ Tesla. Our unit of energy, $\Delta$, is 
the broadening of the localized QD levels, i.e., $\Delta = \pi V^2 \rho \left( \epsilon_{F}\right)$, where 
$\rho \left( \epsilon_{F}\right)$ is the density of states at the Fermi energy. 

\begin{figure}[h!]
\center
\includegraphics[width=\columnwidth]{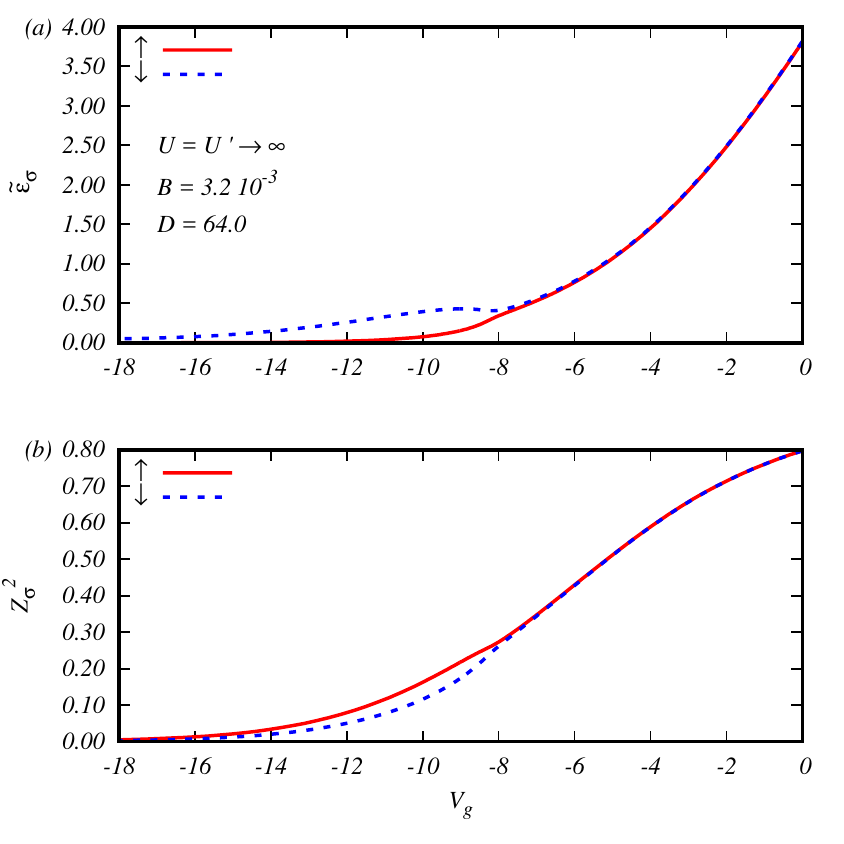}
	\caption{(Color online)(a) Renormalized energy $\tilde{\epsilon}_{\sigma}$ and (b) $Z^2_{\sigma}$ 
	as a function of the gate potential $V_g$ 
	for $\sigma=\uparrow$ [solid (red) curve] and $\sigma=\downarrow$ [dashed (blue) curve]
	for the DQD system, with 
	$U = U^{\prime}\rightarrow \infty$, $D=64.0$ and $B=3.2 \times 10^{-3}$.}
\label{figure2}
\end{figure}

We discuss first the results obtained using the MFSBA. The renormalized spin dependent QD local energy $\tilde{\epsilon}_{\sigma}$, shown in Fig.~\ref{figure2}(a) as a function of the gate potential, is the same for both QDs, but is nevertheless spin dependent due to the applied external magnetic field. 
Results for $\sigma=\uparrow$ and $\sigma=\downarrow$ are given by the solid (red) and the dashed (blue) curves, respectively. 
As $V_g$ decreases, starting around the Fermi energy $\epsilon_{F}=0$, the renormalized energies (for different spin projections) are undistinguishable down to $V_g \approx -8$, where they split ($\tilde{\epsilon}_{\downarrow} > \tilde{\epsilon}_{\uparrow}$). This indicates that a change in the ground state occurs for $V_g \lesssim -8$, region in the parameter space where the ground state SU(4) symmetry is lost. 
In particular, continuously reducing $V_g$, the renormalized energy $\tilde{\epsilon}_{\uparrow}$ displays a typical Kondo behavior, within the MFSBA approach, being almost independent of the gate potential and taking a value in the immediate vicinity of the Fermi energy, representing the $SU(2)$ Kondo peak, while $\tilde{\epsilon}_{\downarrow}$ maintains its value above the Fermi energy. 

\begin{figure}[h!]
\center
\includegraphics[width=\columnwidth]{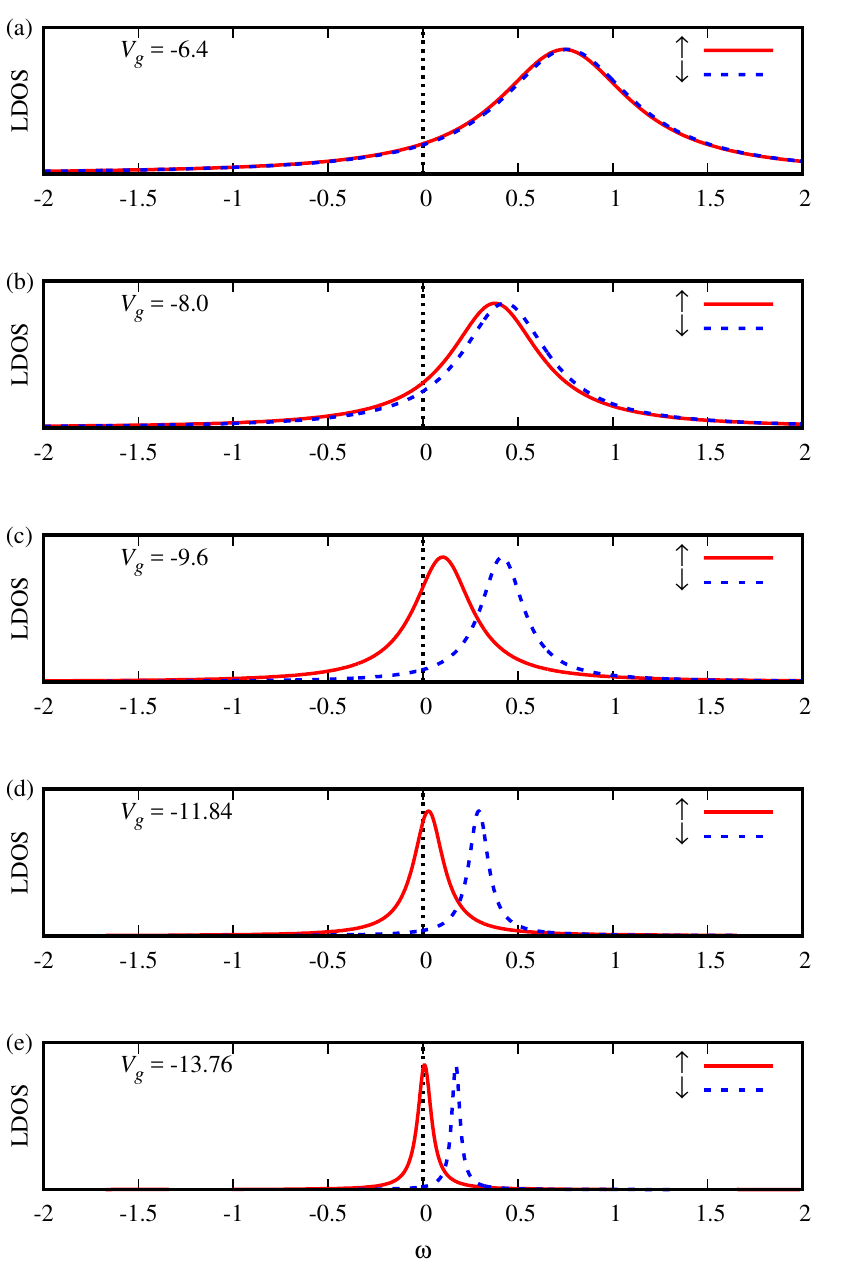}
\caption{(Color online) Local density of states as a function of $\omega$ for $\sigma=\uparrow$ 
	[solid (red) curve] and $\sigma=\downarrow$ [dashed (blue) curve], for 
	$U = U^{\prime} \rightarrow\infty$, $D=64.0$,  $B=3.2 \times 10^{-3}$, and different values of 
	$V_g= -6.4$ (a), $-8.0$ (b), $-9.6$ (c), $-11.84$ (d) and $-13.76$ (e).}
\label{figure3}
\end{figure}

This spin-dependent splitting also occurs for the parameter $Z^2_{\sigma}$, which renormalizes the matrix elements that connect the QDs to the electron reservoirs $\tilde{V}_{\sigma}=VZ_\sigma$, as shown in Fig.~\ref{figure2}(b), where $Z^2_\sigma$ decreases with the gate potential, and takes different values for different spin orientations for $V_g < -8.0$, in agreement with Fig.~\ref{figure2}(a). 
As $\tilde{V}_{\sigma}$ controls the width of the peak associated to $\tilde{\epsilon}_{\sigma}$, one expects that the peak for $\sigma=\uparrow$, which reaches the Fermi level ($\epsilon_F=0$) as $V_g$ decreases [solid (red) curve in Fig.~\ref{figure2}(b)], and therefore determines the properties of the Kondo ground state, such as the Kondo temperature, will get narrower as the (${SU(4)}-{SU(2)}$ transition occurs, implying that $T_K^{SU(4)}>>T_K^{SU(2)}$. 
This will be shown to be indeed the case by an explicit calculation of the width of the QD levels, as shown next, in Fig.~\ref{figure3}. 

The results shown in Figs.~\ref{figure2}(a) and (b) can be better understood by comparing the QD's local density of states (LDOS), for each spin projection, for gate potential values above and below $V_g = -8$, where the $\tilde{\epsilon}_{\sigma}$ splitting occurs. The LDOS results for the two identical  QDs are shown in Fig.~\ref{figure3}, for $V_g=-6.4$ [panel (a)], $-8.0$ [(b)], $-9.6$ [(c)], $-11.84$ [(d)], and $-13.76$ [(e)], for $\sigma=\uparrow$ [solid (red) curves] and $\sigma=\downarrow$ [dashed (blue) curves]. Figure \ref{figure3}(a) illustrates the situation for gate potential values above the splitting, where the LDOS peaks for both spin projections are essentially superposed, showing that although the magnetic field has broken the $SU(4)$ symmetry, the ground state preserves it, as this better minimizes its energy.  Although not explicitly shown, this situation 
prevails in the interval $-8.0 < V_g < 0$. 
As $V_g$ keeps decreasing, the LDOS peak narrows and splits up, both of the resulting peaks still located above the Fermi energy, as shown in panel (b) of Fig.~\ref{figure3}. 
Therefore, below $V_g = -8$, the ground state responds to the Zeeman splitting, caused by the magnetic field, by explicitly taking the Hamiltonian\textquotesingle s $SU(2)$ symmetry, as now this better minimizes its energy. 
This $SU(2)$-Kondo is an \emph{orbital}-Kondo state, its degenerate DQD states being (using notation from Table \ref{table:slavebosons}) $|0;\uparrow\rangle$ and $|\uparrow;0\rangle$. 

Further decreasing $V_g$ leads to further narrowing of both peaks, accompanied by a larger splitting between them, which is achieved by the $\sigma=\uparrow$ peak accelerating its shift towards the Fermi energy, while the $\sigma=\downarrow$ peak moves slightly up in energy. 
The narrowing of the peaks, as first discussed in relation to the variation of $Z^2_\sigma$ with $V_g$ [see Fig.~\ref{figure2}(b)], is compatible with the fact that the Kondo temperatures of the $SU(2)$ and $SU(4)$ Kondo ground states satisfy $T_K^{SU(4)}>>T_K^{SU(2)}$ (see Ref.~\onlinecite{Lim2006}). 
This is clearly illustrated by the sizable narrowing of the solid (red) peak from panel (a) to panel (e) in Fig.~\ref{figure3}. 
 
As will be discussed below in detail, the spin-dependent renormalization reflects the high spin filter efficiency of the device and it is also critical to understand, within the MFSBA, the abrupt changes in the QD's occupation as a function of the gate potential.

\begin{figure}[h!]
\includegraphics[width=\columnwidth]{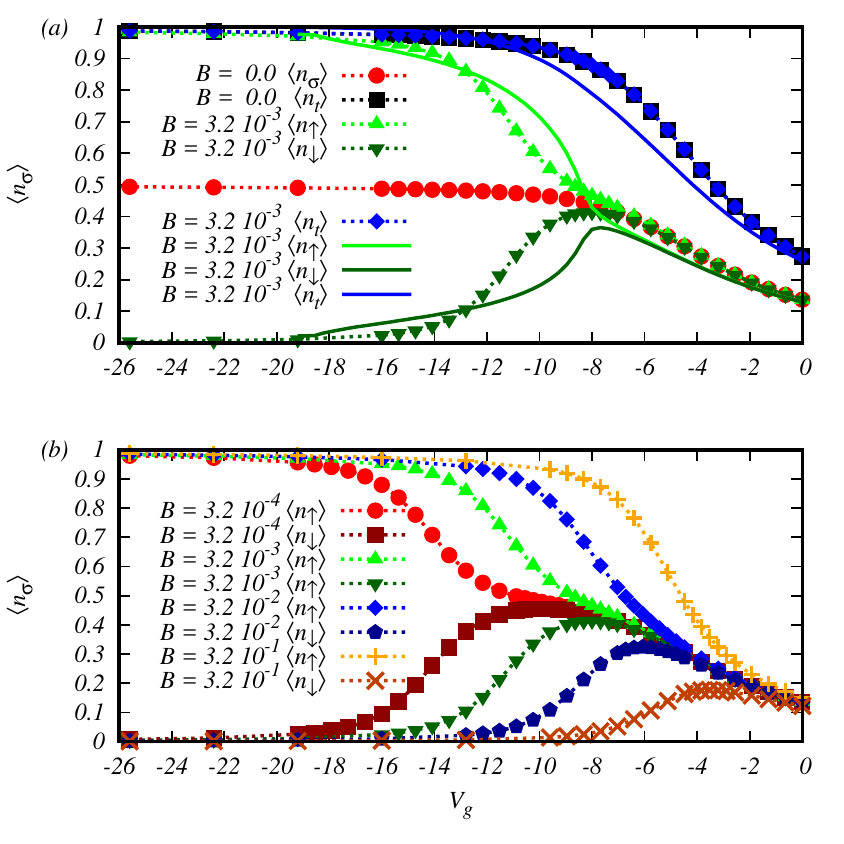}
	\caption{(Color online) QD occupation numbers $\langle n_{\sigma} \rangle$ and $\langle n_{t} \rangle 
	=\langle n_{\uparrow}\rangle + \langle n_{\downarrow} \rangle$ as a function of 
	$V_g$. 
	(a) POA results at zero magnetic field for $\langle n_{\sigma} \rangle$ [(red) circles] 
	and $\langle n_{t} \rangle$ [(black) squares], as well as 
	a comparison of $\langle n_{\uparrow} \rangle$, $\langle n_{\downarrow} \rangle$, and 
	$\langle n_{t} \rangle $ 
	results obtained with POA (symbols) with those obtained with MFSBA (solid lines), 
	for $B=3.2 \times 10^{-3}$. Note that results for MFSBA 
	and POA  agree semi-quantitatively. (b) POA $\langle n_{\sigma} \rangle$ 
	results for $B=3.2 \times 10^{-4}$, $3.2 \times 10^{-3}$, $3.2 \times 10^{-2}$, and $3.2 \times 10^{-1}$. 
	All results in both panels are for $D=64.0$,}
\label{figure4}
\end{figure}

Taking the same parameters as in Figs.~\ref{figure2} and \ref{figure3}, the spin dependent electron 
occupation in each QD $\langle n_{\sigma} \rangle$, as a function of $V_g$, is calculated 
using POA and MFSBA, as shown in Figs. \ref{figure4}(a) and (b). In the case of MFSBA, the 
occupation numbers are calculated by integrating the density of states at the QDs obtained from 
the corresponding Green's function. To calculate the same quantities in the POA formalism, we 
take the derivative of the ground state energy with respect to the gate potential $V_g$. 
The $B=3.2 \times 10^{-3}$ results in Fig.~\ref{figure4}(a) show a semi-quantitative agreement between 
POA (symbols) and MFSBA (solid lines). 

Inspecting the $\langle n_{\sigma} \rangle$ POA results in Fig.~\ref{figure4}(b), for four different Zeeman splitting values, 
$B=3.2 \times 10^{-4}$, $3.2 \times 10^{-3}$, $3.2 \times 10^{-2}$, and $3.2 \times 10^{-1}$, 
shows that while the QD level $V_{g}$ is near the Fermi level and, as a consequence, the QDs are 
still in the charge fluctuating regime, the two smaller Zeeman splittings 
($B=3.2 \times 10^{-4}$ and $3.2 \times 10^{-3}$) are not 
able to minimize the energy of the system (thus polarizing the QDs) when compared to the gain in 
energy brought by the $SU(4)$-Kondo-singlet ground state. 
Therefore, in this regime, the magnetic field is not a relevant quantity, as the ground state does 
not reflect the broken $SU(2)$ symmetry introduced by the field, as already discussed above 
(see also Ref.~\onlinecite{Keller2014}). 
However, as the gate potential is further reduced, and the Kondo temperature $T_K^{SU(4)}$
exponentially decreases, eventually becoming smaller than $B$, a 
sudden change in the behavior of the occupation numbers $\langle n_{\sigma} \rangle$ 
occurs: $\langle n_{\downarrow} \rangle$ reaches a maximum and undergoes a sharp drop, tending to 
zero as $V_g$ is further reduced, while $\langle n_{\uparrow} \rangle$ keeps increasing, 
eventually saturating at $\langle n_{\uparrow} \rangle =1$. Obviously, this occurs because the 
Zeeman splitting $B$ has overtaken $T_K^{SU(4)}$. 
On the other hand, for the larger Zeeman splittings ($B=3.2 \times 10^{-2}$ and $3.2 \times 10^{-1}$), 
the polarization starts to occur for considerably larger values of $V_g$, as 
a small decrease in $V_g$ will be enough to make $T_K^{SU(4)} \lesssim B$. 
It should be clear, however, that the discussion above does not imply that 
$B$ should be compared to the zero-field $T_K^{SU(4)}$, as a finite magnetic 
field does suppress the Kondo temperature, as shown in Fig.~\ref{figure7}(c) \cite{Pablo_TK,Tosi2012}. 

The inflexion point in the function $\langle n_{\downarrow} \rangle \left(V_g\right)$ (where $\nicefrac{d \langle n_{\downarrow} \rangle}{dV_g}=0$) will be used to characterize the $SU(4)-SU(2)$ crossover. 
The results in Fig.~\ref{figure4}(b) indicate that $V_g^{max}$, the value where the maximum for $\langle n_{\downarrow} \rangle$ occurs, as expected, strongly depends upon the magnetic field: for larger $B$ values, the split between $\langle n_{\uparrow} \rangle$ and $\langle n_{\downarrow} \rangle$ occurs for values of $V_g^{max}$ nearer to the Fermi energy. On one hand, this reflects the fact that, as the field increases for a fixed value of gate potential, a Zeeman-split ground state will eventually have a lower energy than an $SU(4)$-Kondo-singlet ground state. On the other hand, the lower is $B$, more charging of the QDs will be required to achieve a splitting, thus resulting in a lower value of $V_g^{max}$. 

\begin{figure}[h!]
\center
\includegraphics[width=\columnwidth]{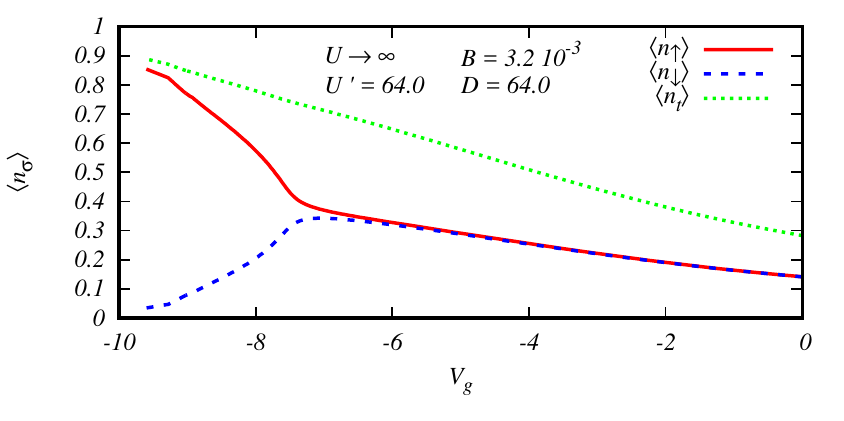}
	\caption{(Color online) MFSBA results for $\langle n_\uparrow \rangle$ [(red) solid],
	$\langle n_\downarrow \rangle$ [(blue) dashed], and $\langle n_{t} \rangle$ [(green) dotted curve], 
	as a function of gate potential in a DQD system, for $D=64.0$, $U \rightarrow \infty$, 
	$U^{\prime} = 64.0$, and $B = 3.2 \times 10^{-3}$. 
	Note that the results are qualitatively the same as the ones obtained for 
	$U = U^{\prime} \rightarrow \infty$ (compare with Fig.~\ref{figure4}).}
\label{figure5}
\end{figure}

At this point, it is interesting to mention that the qua\-li\-tative results for the occupation numbers do not depend upon taking $U^{\prime}\rightarrow\infty$. As the MFSBA calculations are not restricted to the condition $U=U^{\prime}$, we show in Fig.~\ref{figure5} the variation of $\langle n_{\uparrow} \rangle$ and $\langle n_{\downarrow} \rangle$ with $V_g$ for $U^{\prime}=64.0$, keeping $U\rightarrow\infty$. 
The results obtained qualitatively agree with results for $U=U^{\prime}\rightarrow\infty$.
As mentioned above, although in this case the Hamiltonian does not have an explicit $SU(4)$ symmetry (not only because of the pre\-sen\-ce of a finite magnetic field, but also because $U^{\prime} \neq U$), the ground state of the DQD system still preserves this symmetry (up to $V_g \approx -7.0$), as an emergent property \cite{Tosi2013}, and an $SU(4) - SU(2)$ crossover still occurs (compare with Fig.~\ref{figure4}). 

\begin{figure}[h!]
\center
\includegraphics[width=\columnwidth]{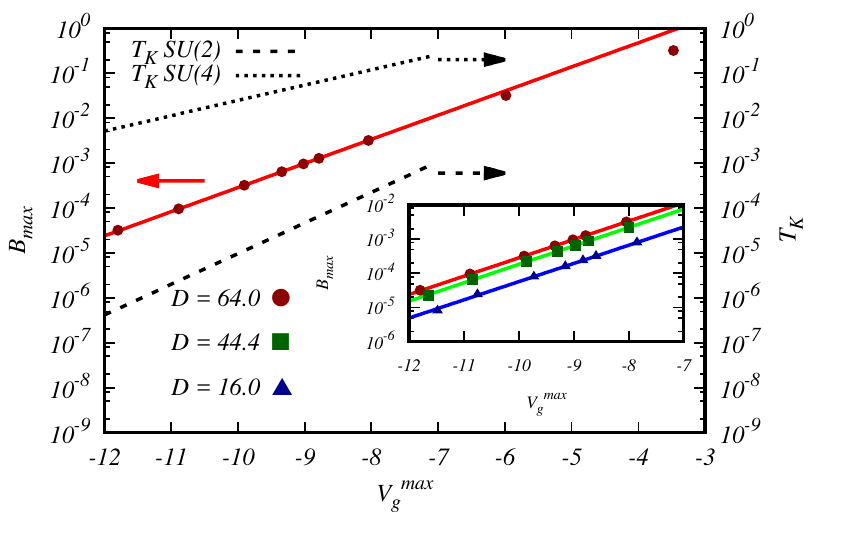}
	\caption{(Color online)  Semi-log plot of the $B_{max}=f\left(V_g^{max}\right)$ function calculated, using POA,  
	for three different values of $D=64.0$ [(red) circles], $44.4$ [(green) squares], and $16.0$ 
	[(blue) triangles]. Note that the lack of $D$-dependence of the exponential factor $a$ 
	(see eq.~\ref{eq:POA B}) reveals the 
	universal character of the $SU(4)-SU(2)$ crossover, which is driven by the Zeeman splitting. 
	The Kondo temperature curves for $SU(4)$ (dotted) and $SU(2)$ (dashed) symmetries 
	(obtained from eq.~\ref{eq:POA Tk}) show that 
	$B_{max}$ has an energy scale intermediate between $T_{K}^{SU(4)}$ and $T_{K}^{SU(2)}$.}
\label{figure6}
\end{figure}

\begin{figure}[h!]
\center
\includegraphics[width=0.95\columnwidth]{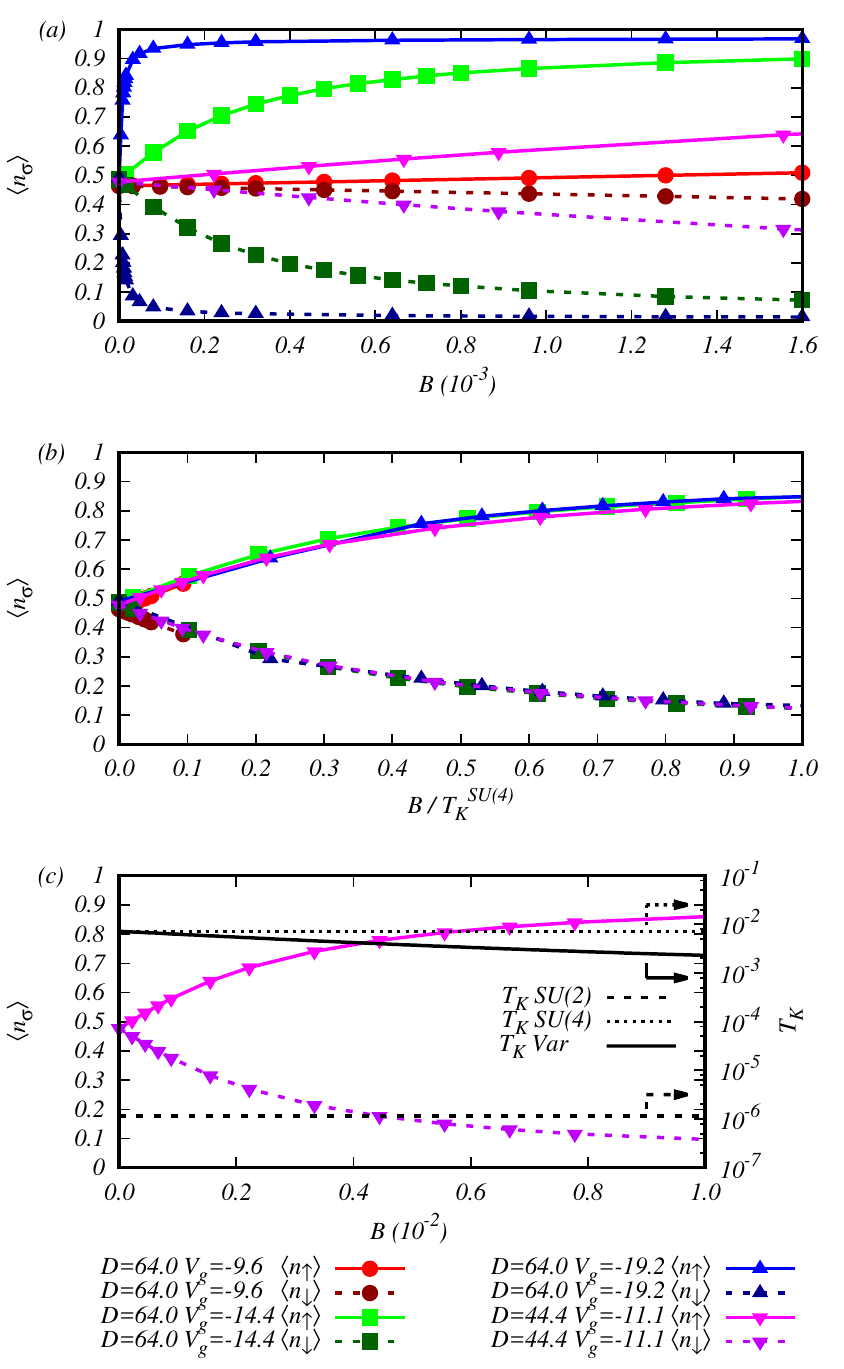}
	\caption{(Color online) (a) POA results for $\langle n_{\uparrow} \rangle$ (solid lines) and 
	$\langle n_{\downarrow} \rangle$ (dashed lines), 
	as a function of Zeeman splitting $B$, for four different values of $V_g=-9.6$ [(red) circles], $-11.1$ 
	[(magenta) down triangles], $-14.4$ [(green) squares], and $-19.2$ 
	[(blue) up triangles]. Note both the very gradual spin polarization 
	when the system is in the charge fluctuation regime [(red) circles], and the very 
	abrupt transition, at very small fields, from $SU(4)$ to $SU(2)$ Kondo 
	when the system is deep into the $SU(4)$ Kondo state at zero field [(blue) up triangles]. 
	(b) Same data as in panel (a), but now plotted against $B/T_K^{SU(4)}$ instead of just $B$. 
	All data sets collapse into two single curves, one for each spin orientation.  
	(c) Left vertical axis: POA results for $\langle n_{\sigma} \rangle$ as a function of Zeeman splitting $B$ for 
	$V_g=-11.11$ [(magenta) down triangles]. Right vertical
	 axis (log scale): Kondo temperatures $T_K^{SU(4)}$ (dotted line) and $T_K^{SU(2)}$ (dashed line) 
	 at $B=0$ are represented by horizontal lines. 
	The solid line is the Kondo temperature for the `crossover state', as a function of $B$,  obtained as a 
	variational interpolation between the $SU(4)$ and $SU(2)$ states \cite{Pablo_TK,Tosi2012,note-variational}. 
	All calculations done for $D=64.0$, except for the $V_g=-11.1$ results, which were obtained for $D=44.4$. }
\label{figure7}
\end{figure}

It is believed that, in the presence of a magnetic field, a broken $SU(4)$ symmetry will be 
clearly observable only when $B \approx T_{K}^{SU(4)}$\cite{Pablo_TK}. 
In order to clarify this point, in 
Fig.~\ref{figure6} we present a semi-log plot with POA results for the Zeeman splitting $B_{max}$ 
in the left axis (in logarithmic scale), and the corresponding values of 
$V_g^{max}$, at which the maximum in $\langle n_{\downarrow} \rangle$ occurs, in the horizontal axis. 
The variation in $B_{max}$ spans more than four orders of magnitude. The main panel results are for 
$D=64.0$ [(red) circles], with two extra sets of results plotted in the inset, for $D=44.4$ [(green) squares] 
and $16.0$ [(blue) triangles]. The results in the main panel and in the inset clearly show an exponential dependence of 
$B_{max}$ on $V_g^{max}$, therefore a least squares fitting was done, using the expression 
\begin{equation}
\label{eq:POA B}
	B_{max}=D\exp(aV_g^{max}),   
\end{equation}
and the results of these fittings were plotted as solid lines. 
The value of the Zeeman splitting, $ B_{max} $, is the relevant energy scale that controls the $SU(4)-SU(2)$ crossover, 
which, according to our definition, occurs when $V_g=V_g^{max}$. 
This energy scale has a universal behavior in the Kondo regime, as described by eq.~\ref{eq:POA B}, 
extending into the charge fluctuating regime as well, although it looses 
its universal character in the neighborhood of the Fermi energy. This is illustrated in Fig.~\ref{figure6} by 
the fact that the two nearest points to the Fermi energy no longer coincide with the straight line given 
by eq.~\ref{eq:POA B}. The loss of universality is an expected result, clearly showing that the universal 
behavior is restricted to the Kondo regime, as it is the case for the Kondo temperature. Anyhow, it 
is important to emphasize that, for larger values of $B$, as illustrated for $B=3.2 \times 10^{-1}$ in Fig.~\ref{figure4}(b), 
$\langle n_{\downarrow} \rangle$ reaches a maximum along the entire charge 
fluctuation region, therefore defining the energy scale $B_{max}$ as controlling the $SU(4)-SU(2)$ crossover also in this regime.

In addition, the results in the inset for three different values of $D$ (keeping $\Delta$, our unit of energy, constant) 
clearly show that the parameter $a \sim 1.23$, from eq.~\ref{eq:POA B}, is independent of $D$, llustrating the universality 
of the Zeeman splitting scale of energy that characterizes the $SU(4)-SU(2)$ crossover. Finally, the least squares fitting 
of the POA results (points) using eq.~\ref{eq:POA B} also shows that the choice of the band half-width $D$ as prefactor 
is correct, as the fitting recovers, with good numerical accuracy, the values of $D$ used for 
the POA calculations \cite{note-vgdependence}.

Also shown in the same plot (right axis, in logarithmic scale too) are the Kondo temperatures $T_{K}^{SU(2)}$ (dashed line) 
and $T_{K}^{SU(4)}$ (dotted line) obtained through the expression \cite{Newns1987} 
\begin{equation}
\label{eq:POA Tk}
	T_{K}^{SU(N)}=D\exp(\nicefrac{\pi V_g}{N}), 
\end{equation} 
which was obtained 
through a $U \rightarrow \infty$ variational wave function for the ground state of the system \cite{Fulde}, 
which coincides as well with the mean-field solution of a slave boson formalism (also in the same limit) \cite{Newns1987}. 
These curves are shown in order to facilitate the comparison of their exponential dependence on $V_g$, 
as shown in eq.~\ref{eq:POA Tk}, with the exponential dependence of the Zeeman splitting $B_{max}(V_g^{max})$, 
as described in eq.~\ref{eq:POA B}. These two Kondo temperatures are 
displayed just for values of $V_g^{max}<-7$, which roughly corresponds to the Kondo regime, 
to emphasize that the expression above is not valid in the charge fluctuation regime. 

Surprisingly enough, the Zeeman splitting exponent factor $a \sim 1.23$ in 
eq.~\ref{eq:POA B} has an intermediate value between those 
of the $T_{K}^{SU(4)}$ and $T_{K}^{SU(2)}$ Kondo states (see eq.~\ref{eq:POA Tk}): 
$\nicefrac{\pi}{4}< a < \nicefrac{\pi}{2}$. Moreover, 
a simple inspection of Fig.~\ref{figure6} shows that the value of $B_{max}$ is between one 
to two orders of magnitude less than $T_{K}^{SU(4)}$ and equally greater than $T_{K}^{SU(2)}$, the larger 
difference occurring for larger values, in magnitude, of $V_g$, deep into the Kondo regime. 
As the value of the exponent factor controlling the Zeeman splitting is between those corresponding 
to $T_{K}^{SU(4)}$ and $T_{K}^{SU(2)}$, it is possible, under the effect of small magnetic fields, 
the operation of the DQD system in a regime of high spin polarization 
($\langle n_{\uparrow} \rangle \gg~\langle n_{\downarrow} \rangle$), with important consequences 
for its spin filter performance, as discussed in the next section. 

To properly characterize the $SU(4)-SU(2)$ crossover, it is interesting 
to do the opposite of what was done up to now, i.e., instead of fixing the external field 
and analyzing how $\langle n_{\sigma} \rangle$ depends upon $V_g$, we 
study the variation of $\langle n_{\sigma} \rangle$, at fixed $V_g$, as a function of magnetic field. 
This analysis is done using POA. 
The main idea is to use $V_g$ to place the system, at zero field, either well inside the $SU(4)$ Kondo regime 
or closer to the charge fluctuation region, and then analyze how does the 
application of a magnetic field change the system's properties. 
We study the spin occupation numbers $\langle n_{\uparrow} \rangle$ and $\langle n_{\downarrow} \rangle$,  
which are shown in Fig.~\ref{figure7}(a) (where solid lines indicate $\langle n_{\uparrow} \rangle$ and 
dashed ones $\langle n_{\downarrow} \rangle$) for four different values of gate potential: 
$V_g=-19.2$ [(blue) up triangles] places it well
inside the $SU(4)$ Kondo regime, $V_g=-9.6$ [(red) circles] places the system nearer
to the charge fluctuation regime, while $V_g=-14.4$ [(green) squares] places it halfway between these two. 
These three data sets were obtained for $D=64.0$ and we add a fourth one 
[(magenta) down triangles] at $V_g=-11.1$, with a smaller $D=44.4$, 
to analyse the effect of a different half-bandwidth $D$ on the results obtained, as discussed below.
The results in Fig.~\ref{figure7}(a) indicate that, closer to the charge fluctuation regime frontier,
($V_g=-9.6$ and $-11.1$), and even well inside the Kondo $SU(4)$ regime ($V_g=-14.4$), the spin polarization, as measured 
by $\langle n_{\uparrow}\rangle -\langle n_{\downarrow} \rangle$, is gradually raised in response to 
an increasing (from zero) magnetic field (see the circles, down triangles, and squares curves). 
This behavior can be explained by the larger values of $T_K^{SU(4)}$ for $V_g$ values closer to the Fermi energy 
(see dotted curve in Fig.~\ref{figure6}) as it will take a larger value of field to force the system 
to transition from the $SU(4)$ to the $SU(2)$ regime. This is specially evident for the $V_g=-9.6$ 
results [(red) circles, with the highest $T_K^{SU(4)}$], where a larger field is needed to 
generate a sizable spin polarization. One would expect then that the system will require just 
a very small magnetic field to transition from the $SU(4)$ Kondo regime to the 
orbital $SU(2)$ Kondo regime once $T_K^{SU(4)}$ decreases substantially. This is exactly what is observed for 
$V_g=-19.2$ [(blue) up triangles], where $T_K^{SU(4)}$ is much smaller 
(see Fig.~\ref{figure6}) and the system responds much more abruptly to the magnetic field. 
In reality, even results for $V_g=-14.4$ [(green) squares], where $T_K^{SU(4)}$ is not so low, show that a small external 
magnetic field $h \approx 0.1$ Tesla (corresponding to $B \approx 0.0022$, if one takes, for instance, the 
gyromagnetic factor for $GaAs$), is enough to obtain a sizable spin polarization, as shown in Fig.~\ref{figure7}(a). 

The results in Fig.~\ref{figure7}(a), despite being interesting, were somewhat expected. 
What makes them more relevant are the results presented in panel (b), where it is shown that if the 
$\langle n_{\uparrow} \rangle$ and $\langle n_{\downarrow} \rangle$ data in panel (a) are plotted 
against $B/T_K^{SU(4)}$ (with $T_K^{SU(4)}$ as obtained from eq.~\ref{eq:POA Tk}), instead of against just $B$, 
all the curves for different parameters collapse into each other. This is true even for the $V_g=-11.1$ data 
[(magenta) down triangles], which has a different value of $D$ in relation to the other data sets. 
This universality result shows that there is a deep connection between the spin polarization 
and the $B/T_K^{SU(4)}$ ratio when an external magnetic field is applied. It is important to 
emphasize that this universality is obtained when adopting 
eq.~\ref{eq:POA Tk} to calculate $T_K^{SU(4)}$, which gives additional support to the use 
of eq.~\ref{eq:POA Tk} to describe the $SU(4)$ Kondo state in the $U \rightarrow \infty$ limit. 

In Fig.~\ref{figure7}(c) we reproduce (left axis) the $\langle n_{\sigma} \rangle$ 
results for $V_g=-11.1$ and $D=44.4$, as a function of Zeeman splitting $B$ [(magenta) down triangles], 
together with (right axis, in log scale) the Kondo temperatures $T_K^{SU(4)}$ (dotted line) 
and $T_K^{SU(2)}$ (dashed line) at zero magnetic field (thus, shown as horizontal lines), obtained 
from eq.~\ref{eq:POA Tk}. 
As previously discussed, in the crossover region the system is in a Kondo ground state that is going 
through a transformation from $SU(4)$ to $SU(2)$ symmetry. 
An estimation of the Kondo temperature of this `crossover state', 
and its dependence on the magnetic field, can be obtained from a variational calculation that interpolates, 
as a function of the magnetic field, between 
$T_K^{SU(4)}$ at $B=0$ and $T_K^{SU(2)}$ obtained for $B \rightarrow \infty$ \cite{note-variational}. 
This interpolated Kondo temperature, denoted as $T_K^{Var}$, is shown in Fig.~\ref{figure7}(c) as a black solid curve. 
Obviously, it starts at $T_K^{SU(4)}$, decreases with $B$, and, for the small interval 
of field variation in the figure, it stays at least three orders of magnitude above $T_K^{SU(2)}$. 
In addition, for $B \approx 0.0022$ (which corresponds to $h \approx 0.1$ Tesla, as mentioned above), for example, 
$T_K^{Var}$ is almost equal to $T_{K}^{SU(4)}$, which, for the parameter values taken, 
results to be of the order of $10K$. These values of field and temperature are perfectly accessible 
experimental conditions for operation of the DQD as a spin-filter, as described in the next section. 

\section{Spin Filter}\label{sec5}

Besides the natural intrinsic interest in systems whose properties depend on spin orientation, they are also important because, under adequate control, they can have very significant applications. The spin filter properties of a QD, or structures of QDs, is one of these very interesting aspects that have been studied in the last years\cite{Recher2000,Borda2003,Feinberg2004,Hanson2004,Dahlhaus2010,Mireles2006,Hedin2011}. 
The proposal of producing polarized lead currents as they go through a QD is based on the idea that the Zeeman splitting can be made much stronger in the QD than in the leads, thus creating a spin filter. 
Spin filter phe\-no\-me\-na are obtained when the QD spin-up sublevel is located in the transport window, while the spin-down one can be manipulated to be just outside of it. This requires high magnetic fields (even considering renormalized $g$ factors for the QD) and weak coupling of the QD to the leads, therefore resulting in very sharp localized states, thus properly separating in energy the spin-up from the spin-down level. 
The first restriction introduces experimental limitations to the applicability of the device, while the last condition reduces significantly the intensity of the current circulating through it.
Neither of these difficulties are present in our case because our DQD system, being in the Kondo regime, has a very sharp Kondo spin-polarized level, tuned to be at the vicinity of the Fermi energy, well separated from the other spin polarization [see, for example, Fig.~\ref{figure2}(a)]. 
As the device is required to be in the Kondo regime, the temperature should be below the Kondo temperature, which is a limitation. 
Fortunately, however, the Zeeman splitting required to separate $\langle n_{\uparrow}\rangle$ from $\langle n_{\downarrow}\rangle$, as already discussed, although below $T_{K}^{SU(4)}$, can be taken to be very near it, much larger than $T_{K}^{SU(2)}$.

In order to clarify these points and to show the spin-filter potentialities of our DQD system, we calculate the current as a function of the relevant parameters. The quantum conductance, a dynamical property, can be obtained, within the context of the MFSBA, using the Keldysh formalism \cite{Keldysh_ref}. The current 
through one of the QDs is given by \cite{Land-Butt_ref}, 

\begin{equation}
\label{J_current}
J_{c}=\dfrac{2e}{h}\int_{-\infty}^{\infty}T(\epsilon)[f(\epsilon-\epsilon_{L} )-f(\epsilon-\epsilon_{R})] d\epsilon
\end{equation}
where $T(\epsilon)$ is the transmission, $f(\epsilon)$ the Fermi-Dirac distribution and $\epsilon_{L,R}$ are the Fermi energies of the left and right reservoirs, respectively. For an infinitesimal bias potential (thus 
in the linear regime, where inelastic processes can be neglected \cite{Meir1992}), from eq.~\eqref{J_current} one obtains the 
familiar expression for the conductance 
\begin{equation}
\label{G_conductance}
	{\rm G}=\dfrac{2e^2}{h}T\left(\epsilon_{F}\right), 
\end{equation} 
where the transmission, at the Fermi energy, is given by \cite{Land-Butt_ref},
\begin{align}
\label{T_transmitance}
T(\epsilon_{F} )=4\pi^{2}{V_e}^{4}\rho_{1}(\epsilon_{F}) \rho_{\bar{1}}(\epsilon_{F}) \vert G_{00}^{\sigma}\left(\epsilon_{F}\right)\vert ^2, 
\end{align}
where $\rho_{1}(\epsilon_{F}) = \rho_{\bar{1}}(\epsilon_{F})$ is the LDOS at the first site of the leads, (see labeling in Fig.~\ref{figure1}). 
For an embedded QD configuration [see Fig.~\ref{figure1}(a)], the Green's function $G_{00}^{\sigma}\left(\epsilon _{F}\right)$ is given by $G_{dd}^{\sigma}\left(\epsilon _{F}\right)$, which is the dressed Green's function at the QD, and $V_e=V$. 
In the case of side-coupled QDs [Fig.~\ref{figure1}(b)], $V_e$ is the nearest-neighbor hopping matrix e\-le\-ment in the tight-binding representation of the leads, i.e., $V_e = t$, and $G_{00}^{\sigma}\left(\epsilon _{F}\right)$ is given by
\begin{align}
\label{G_00}
G_{00}^{\sigma}(\epsilon _{F})=g_{0} + g^2_{0}V^2G_{dd}^{\sigma}(\epsilon _{F}),
\end{align} 
where $g_{0} = -i/\sqrt{4t^2 - w^2}$ corresponds to the Green's function at the first site of a semi-infinite tight-binding chain. 

This calculation is straightforward for the MFSBA, as the Green's functions can be obtained directly. 
From the perspective of POA, their values at the Fermi energy have to be calculated from the previously obtained electronic occupations at the QDs, using the Friedel sum rule \cite{Hewson}. 
In the next few paragraphs we briefly describe how to do that.

The Green's function for a QD connected to an electron reservoir can be written as 
\begin{equation}
G_{dd}^{\sigma}\left(\omega\right) =\frac{1}{\omega -V_{g}-\Sigma _{1B}\left(\omega\right) -\Sigma _{MB}\left(\omega\right) +i\eta},
\end{equation}
where $\Sigma _{1B}\left(\omega\right)$ and $\Sigma _{MB}\left(\omega\right)$ are the one- and many-body self-energies, respectively; and $\eta$ is a small displacement in the imaginary plane to regularize the Green's function for values of $\omega$ outside the band defined by the Fermi sea.

For simplicity, we assume a flat band to describe the leads density of states. 

Using the identity,
\begin{align}
&\frac{\partial}{\partial \omega}\ln \left[ G_{dd}^{\sigma}\left(\omega\right) \right] ^{-1}=\nonumber\\
&G_{dd}^{\sigma}\left(\omega\right)\left(1-\dfrac{\partial}{\partial \omega} \Sigma _{1B}\left(\omega\right) -\dfrac{\partial}{\partial \omega}\Sigma _{MB}\left(\omega\right)\right),
\end{align}
then integrating both sides, using that 
\begin{equation}
\langle n_{\sigma}\rangle =-\dfrac{2}{\pi}\int _{-\infty}^{\epsilon _{F}}\Im \left\{ G_{dd}^{\sigma}\left(\omega\right)\right \} d\omega
\end{equation}
(where $\Im \left\{\cdots \right \}$ means taking the imaginary part) and imposing 
the Fermi liquid conditions\cite{Hewson}, we obtain that 
\begin{align}
&\Im \left\{-\dfrac{1}{\pi}\ln\left[\left(G_{dd}^{\sigma}\left(\omega\right)\right) ^{-1}\right]\right\}_{-\infty}^{\epsilon _{F}}
=\frac{\langle n_{\sigma}\rangle}{2}.
\label{eq:POA_C_rel1}
\end{align}

Now, we explicitly introduce the phase of the Green's function, 
\begin{equation}
G_{dd}^{\sigma}\left(\omega\right) =\vert G_{dd}^{\sigma}\left(\omega\right)\vert e^{i\phi\left(\omega\right)}.
\end{equation}
The asymptotic behavior of the one-body propagator, 
$G_{dd}^{\sigma}\left(\omega\rightarrow \infty\right)=\nicefrac{1}{\left(\omega +i\eta\right)}$, and some 
algebra, allows us to write that  
\begin{equation}
\phi\left(-\infty\right) =\pi,
\end{equation}
and
\begin{equation}
\phi\left(\epsilon _{F}\right)=\pi\left(1-\frac{\langle n_{\sigma}\rangle}{2} \right) .
\label{eq:POA_C_rel2}
\end{equation}

Then, from the definition of $\phi$ and eq.~\eqref{eq:POA_C_rel1}, it is possible to obtain 

\begin{equation}
\vert G_{dd}^{\sigma}\left(\epsilon _{F}\right) \vert ^{2}=\dfrac{\sin ^{2}\left[\frac{\pi}{2}\langle n_{\sigma}\rangle \right]}{\Delta ^{2}}.
\label{eq:POA_G00_n}
\end{equation} 

From eqs.~\eqref{G_conductance}, \eqref{T_transmitance}, and \eqref{eq:POA_G00_n} the conductance can be written in terms of the occupations numbers $\langle n_{\sigma}\rangle$, for the case of the embedded QDs, resulting in
\begin{align}
{\rm G}_{\sigma}\left( \frac{e^{2}}{h}\right) =\sin ^{2}\left[\frac{\pi}{2}\langle n_{\sigma}\rangle\right].
\end{align}

For side-coupled QDs it is possible to relate $\vert G_{00}^{\sigma}\left(\epsilon _{F}\right) \vert ^{2}$ with the electronic ocuppations at the QDs $\langle n_{\sigma}\rangle$ through eq. \eqref{G_00}. Reasoning in an analogous way as just done above, the conductance results to be 
\begin{align}
	{\rm G}_{\sigma}\left( \frac{e^{2}}{h}\right) =1-\sin ^{2}\left[\frac{\pi}{2}\langle n_{\sigma}\rangle \right].
\end{align}

\begin{figure}[h!]
\center
\includegraphics[width=\columnwidth]{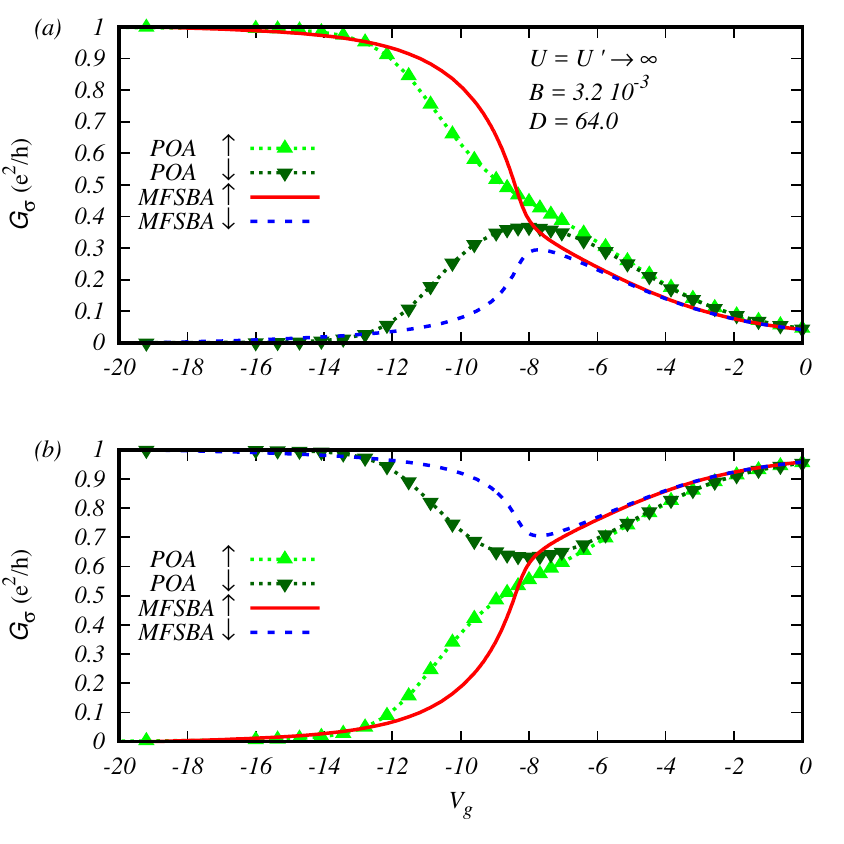}
	\caption{(Color online) MFSBA (lines) and POA (symbols) conductance for spin $\sigma$ electrons versus 
	$V_g$ for $D=64.0$, $B=3.2 \times 10^{-3}$, and $U=U^{\prime} \rightarrow \infty$ 
	for cases (a) embedded and (b) side-coupled to the leads.}
\label{figure8}
\end{figure}

Using the equations just obtained, we show in Fig.~\ref{figure8}(a) MFSBA (lines) and POA (symbols) conductance results obtained for the case of embedded QDs, under the effect of an external magnetic field, as a function of $V_g$. 
An inspection of the figure allows us to conclude that both approaches provide qualitatively equivalent results for the transport properties.  
In the region $V_g < -12.0$ (for both panels) the spin-up conductance is almost $2e^2/h$, while it is close to zero for spin-down. 
This is an interesting result, showing that even for relatively low magnetic fields $B=3.2 \times 10^{-3}$ ($h<0.1$ Tesla, for the case of GaAs), in the appropriate region of the parameter space, the DQD device operates as a very effective spin filter.
It is interesting to notice that, in the case of side-coupled QDs, Fig.~\ref{figure8}(b), the role of the electron spin is interchanged, i.e., the transmitted electrons are down-spins (opposing the field direction), while for embedded QDs the transmitted electrons are up-spins (along the field direction). 
For the side-coupled QD configuration [Fig.~\ref{figure1}(a)], when the system is in a Kondo regime, an up-spin electron circulating through the system has two channels to go through, one connecting the leads directly, and another channel that visits the side-coupled QD. 
As they have opposite phases, the destructive interference between them gives rise to a typical Fano anti-resonance. 
This destructive interference, regarding spin polarization, results in the opposite effect (polarization opposite to field direction) in comparison to embedded QDs. In this case, the spin down electron is the one that is transmitted, while the spin-up conductance rapidly vanishes for decreasing $V_g$, as  shown in Fig.~\ref{figure8}(b). 

\section{Conclusions}\label{sec6}

We studied the $SU(4)-SU(2)$ crossover driven by an external magnetic field for 
two capacitively coupled QDs connected to metallic leads. 
The crossover is characterized by the Zeeman splitting $B_{max}$ at which the $\langle n_{\downarrow}\rangle$ 
has a well-defined maximum as a function of the gate potential for a value denoted as $V_g^{max}$. 
The functional dependence of $B_{max} = f(V_g^{max})$, turns out to have a universal character, 
$B_{max} = D \exp \left( a V_g^{max} \right)$, 
in the Kondo regime, as discussed in detail in Fig~\ref{figure6}. 
This universality is lost as one enters into the charge fluctuating regime, 
the same way as it happens to the Kondo temperature.   
However, it is important to emphasize that the occurrence of the maximum extends into
the valence fluctuating regime, what permits to define the energy scale $B_{max}$ as the magnitude that  controls the $SU(4)-SU(2)$ crossover independently of the system regime. 

We were able to show that already in the crossover region, in an $SU(2)$ ground state, for 
an effective Kondo temperature near the $SU(4)$ one, the electronic populations at the QDs 
are significantly spin polarized along the magnetic field. Moreover, depending upon the parameters 
of the system, this can be obtained even for small magnetic fields ($h \lesssim 0.1$ Tesla 
for the case of GaAs and a Kondo temperature that could be of the order of several degrees Kelvin). 
In that respect, we should mention that, in comparison to a similar device proposed 
in Ref.~\cite{Borda2003}, our device can operate at considerably lower field.  

In addition, this DQD structure was studied adopting the MFSBA and a POA formalisms, which were able to 
describe the mentioned properties, giving qualitatively equivalent results. With this purpose, 
it was necessary to extend the POA, originally derived to study one Kondo impurity, to the 
analysis of two capacitively coupled local levels. 
This extension provides almost exact results, as far as the static 
zero-temperature properties are concerned. 

We conclude that this DQD system, under the influence of a magnetic field, has very interesting 
cross-over properties and, studying its conductance, that it could also operate as an effective 
spin-filter, with potential applications in spintronics. 

\section{Acknowledgment} 
V.L. and R.A.P. acknowledge a PhD studentship from the brazilian agency Conselho Nacional de Desenvolvimento Cient\'{\i}fico e Tecnol\'ogico (CNPq) and E.V.A. acknowledges the financial support from (CNPq) and the brazilian agency Funda\c{c}\~ao de Amparo a Pesquisa e Desenvolvimento do Estado do Rio de Janeiro (FAPERJ).

\appendix

\section{The Mean Field Slave Bosons Approximation}

\begin{figure}[h!]
\center
\includegraphics[width=\columnwidth]{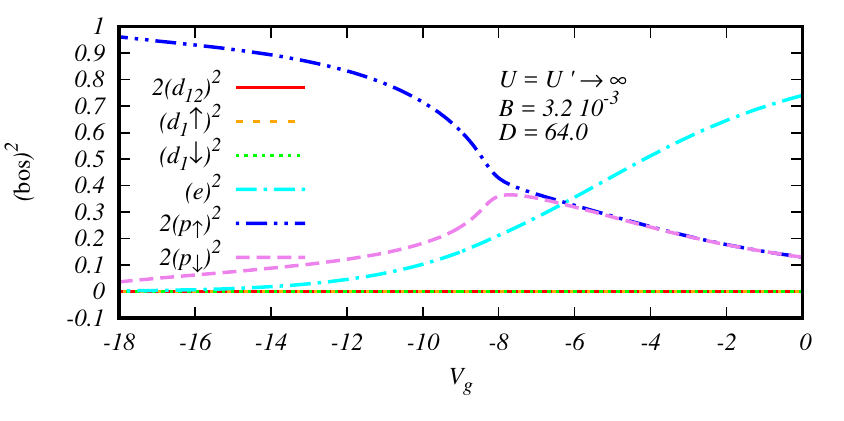}
	\caption{(Color online) Expectation values of the bosonic operators 
	$e^2$, $2p^2_\sigma$, $2d^2_{12}$, $d^2_{1\sigma}$ (per spin $\sigma$)  as a function 
	of $V_g$, for $D=64.0$, $B=3.2 \times 10^{-3}$, and $U=U^{\prime} \rightarrow \infty$.}
\label{figure9}
\end{figure}

In the slave bosons approximation, extra bosonic operators are introduced to represent all the possible states of charge occupation of our DQD system. 
In our case these operators are defined in Table~\ref{table:slavebosons} in the main text. 
The charge conservation condition for each QD and the completeness condition impose relations that the boson operators should fulfill, given by 
\begin{align}
\label{Q}
Q_{j\sigma} =p^{\dagger}_{j\sigma}p_{j\sigma} + d^{\sigma\bar{\sigma}\dagger}_{12}d^{\sigma\bar{\sigma}}_{12}\delta_{1j} 
	+ d^{\bar{\sigma}\sigma\dagger}_{12}d^{\bar{\sigma}\sigma}_{12}\delta_{2j} \nonumber\\ 
	+ d^{1\dagger}_{\sigma}d^{1}_{\sigma} = c^{\dagger}_{dj,\sigma}c_{dj,\sigma},
\end{align}
and
\begin{align}
\label{I}
I=e^{\dagger}e + \sum_{j,\sigma}p^{\dagger}_{j\sigma}p_{j\sigma}+\sum_{\sigma}d^{\sigma\bar{\sigma}\dagger}_{12}
	d^{\sigma\bar{\sigma}}_{12} \nonumber\\ +\sum_{\sigma}d^{1\dagger}_{\sigma}d^{1}_{\sigma} = 1,
\end{align}
where $Q_{j\sigma}$ is the charge per spin in QD $j = 1,2$, for $\sigma = \uparrow/\downarrow$, $I=1$ defines the completeness condition, and $\delta_{ij}$ is the Kronecker delta. 
The fermionic operators of the impurity, in the context of the slave bosons formalism, transform as follows: $c^{\dagger}_{dj,\sigma}$\textrightarrow $Z^{\dagger}_{j,\sigma}c^{\dagger}_{dj,\sigma}$, where the $Z_{j\sigma}$ operator, consisting of all bosonic operators associated with processes in which an electron with spin $\sigma$ is annihilated, is defined as 
\begin{align}
\label{Z}
Z_{j\sigma}=Q^{-\frac{1}{2}}_{j\sigma}(e^{\dagger}p_{j\sigma} + p^{\dagger}_{\bar{j}\bar{\sigma}}(d^{\sigma\bar{\sigma}}_{12}\delta_{1j} + d^{\bar{\sigma}\sigma}_{12}\delta_{2j}) + \nonumber\\ p^{\dagger}_{\bar{j}\sigma}d^{1}_{\sigma})(1 -Q^{-\frac{1}{2}}_{j\sigma}). 
\end{align} 
The mean field approximation of this formalism, the so-called MFSBA, consists in replacing the bosonic operators by their mean values. 
For the sake of simplicity, they are named by the same letter as the operators themselves. 
These mean values and the Lagrange multipliers $\lambda$ and $\lambda_{\sigma}$, incorporated to satisfy the slave boson conditions, are determined by minimizing the free energy of the system. 
These conditions create a set of nine non-linear equations (one for each of the six bosonic operators and three Lagrange multipliers), which should be self-consistently solved to obtain the parameters of the effective one-body Hamiltonian: 

\begin{flalign}
\frac{\partial\langle H_{eff}\rangle}{\partial e} &= 2\sum_{\sigma}V\frac{\partial Z_\sigma}{\partial e}\left(\langle c^{\dagger}_{k,\sigma}c_{d,\sigma}\rangle + h.c\right) + 2\lambda e = 0, &\\
\frac{\partial\langle H_{eff}\rangle}{\partial p_\uparrow} &= 2\sum_{\sigma}V\frac{\partial Z_\sigma}{\partial p_\uparrow}\left(\langle c^{\dagger}_{k,\sigma}c_{d,\sigma}\rangle + h.c\right) \nonumber\\ & \quad + 4(\lambda - \lambda_{\uparrow}) p_\uparrow = 0, \\
\frac{\partial\langle H_{eff}\rangle}{\partial p_\downarrow} &= 2\sum_{\sigma}V\frac{\partial Z_\sigma}{\partial p_\downarrow}\left(\langle c^{\dagger}_{k,\sigma}c_{d,\sigma}\rangle + h.c\right) \nonumber\\ & \quad + 4(\lambda - \lambda_\downarrow) p_\downarrow = 0, \\
\frac{\partial\langle H_{eff}\rangle}{\partial d_{12}} &= 2\sum_{\sigma}V\frac{\partial Z_\sigma}{\partial d_{12}}\left(\langle c^{\dagger}_{k,\sigma}c_{d,\sigma}\rangle + h.c\right) \nonumber\\ & \quad +4(\lambda -\lambda_\uparrow - \lambda_\downarrow + U^{\prime}) d_{12} = 0, \\
\frac{\partial\langle H_{eff}\rangle}{\partial d_{1\uparrow}} &= 2\sum_{\sigma}V\frac{\partial Z_\sigma}{\partial d_{1\uparrow}}\left(\langle c^{\dagger}_{k,\sigma}c_{d,\sigma}\rangle + h.c\right) \nonumber\\ & \quad + 2(\lambda -2\lambda_\uparrow + U^{\prime}- 2\mu_BB) d_{1\uparrow} = 0, \\
\frac{\partial\langle H_{eff}\rangle}{\partial d_{1\downarrow}} &= 2\sum_{\sigma}V\frac{\partial Z_\sigma}{\partial d_{1\downarrow}}\left(\langle c^{\dagger}_{k,\sigma}c_{d,\sigma}\rangle + h.c\right) \nonumber\\ & \quad 
+ 2(\lambda -2\lambda_\downarrow + U^{\prime}+ 2\mu_BB) d_{1\downarrow} = 0, \\
\frac{\partial\langle H_{eff}\rangle}{\partial\lambda} &= e^{2} + 2p^{2}_\uparrow + 2p^{2}_\downarrow + 2d^{2}_{12} + d^{2}_{1\uparrow} + d^{2}_{1\downarrow} -1 = 0,  
\end{flalign}

\begin{align}
\frac{\partial\langle H_{eff}\rangle}{\partial\lambda_\uparrow} &= \langle c^{\dagger}_{d\uparrow} c_{d\uparrow} \rangle - p^{2}_\uparrow - d^{2}_{12} - d^{2}_{1\uparrow} = 0, \\
\frac{\partial\langle H_{eff}\rangle}{\partial\lambda_\downarrow}&= \langle c^{\dagger}_{d\downarrow} c_{d\downarrow} \rangle - p^{2}_\downarrow - d^{2}_{12} - d^{2}_{1\downarrow} = 0, 
\end{align}
where $H_{eff}$ is given by eq.~\eqref{eq:app_H_eff}, and $e^2$, $p^2_\sigma$, $d^2_{12}$, $d^2_{1\sigma}$, as previously mentioned, are taken to be the mean values of the corresponding bosonic operators. Fig.~\ref{figure9} shows results of all these mean values, as functions of $V_g$, for $U = U^{\prime}\rightarrow\infty$ and $B = 10^{-4}$. 
For positive values of $V_g$, the empty QD state, represented by the meanvalue $e^2$, is dominant, but rapidly decreases as $V_g$ approaches the Fermi level. 
We can observe the splitting of the spin dependent occupancy $p_\sigma^2$, for $V_g \approx -4.0$, indicating the $SU(4)-SU(2)$ crossover. 
The double occupancy state $|\uparrow,\uparrow\rangle$ has probability $d^2_{1\uparrow}=0$, as it costs an infinite energy to simultaneously populate the QDs with two electrons due to the infinite $U^{\prime}$ inter-dot Coulomb repulsion. 
For a finite value of $U^{\prime}$, the occupation numbers (not shown), in the parameter region $V_g > -U^{\prime}$, are almost identical to those for $U^{\prime}=U\rightarrow\infty$. 
This indicates that in this region of parameter space the value of $U^{\prime}$ does not change the results qualitatively.

\section{The Projection Operator Approach}

As discussed in the main text, the central idea of the POA is to separate the Hilbert space of the system of interest, which ground state $\vert \Psi\rangle$ obeys,
\begin{equation}
H\vert \Psi\rangle=E\vert \Psi\rangle ,
\end{equation}
into two different subspaces: (i) the subspace $S_{1}$, containing a single state, denoted $\vert 1\rangle$ and (ii) subspace $S_{2}$, containg the rest of the states in the Hilbert space, which are generically denoted as $\vert 2\rangle$. 
The idea is to choose $\vert 1\rangle$ so that, by operating in $S_{1}$ with a renormalized Hamiltonian, one can obtain not only the ground state energy $E$, but also some of its static properties \cite{POA_1,POA_2}. 
The renormalized Hamiltonian that operates in the $S_{1}$ subspace can be written as,
\begin{equation}
H_{\rm ren} = H_{11}+H_{12}\left( E-H_{22}\right)^{-1}H_{21} ,
\end{equation}
where, 
\begin{equation}
H_{ij}=\vert i\rangle\langle i\vert H\vert j\rangle\langle j\vert,
\end{equation}
such that the renormalized Hamiltonian satisfies,
\begin{equation}
\label{eq:E_appA}
H_{\rm ren}\vert 1\rangle = E \vert 1\rangle
\end{equation}
that permits trivially to obtain, 
\begin{equation}
\label{eq:E_appB}
\langle 1\vert H_{\rm ren}\vert 1\rangle = E.
\end{equation}

The self-consistent solution of this last equation, -the renormalized Hamiltonian depends explicitly upon the energy $E$-, permits to find the ground state energy $E$ of the system. It is important to adequately choose the state $\vert 1\rangle$. We take it as given by the ground state of the two Fermi seas and the two uncharged QDs. 
All other states that belong to subspace $S_{2}$ can be obtained by successive applications of the Hamiltonian $H_{21}$ on state $\vert 1\rangle$.

To obtain the ground state energy it is necessary to calculate $\langle 1\vert H_{\rm ren}\vert 1\rangle$. The first term is the expected value of $H_{11}$, given by,
\begin{equation}
\epsilon_{T}=\langle 1\vert H_{11}\vert 1\rangle =2\sum_{\epsilon_{\pmb{k}}<\epsilon_{F}}\epsilon_{\pmb{k}}
\end{equation}

The contribution to the energy of subspace $S_{2}$ is calculated assuming the QDs to be connected to identical leads through matrix elements $V_{\pmb{k_{j}}}=V$ that are taken to be independent of the momentum $\pmb{k_{j}}$. 
The energy can be written as\cite{POA_1,POA_2},

\begin{equation}
E=\Delta E+2\epsilon_{T}
\end{equation}

\begin{equation}
\Delta E=f_{1}\left(\Delta E\right)
\end{equation}

\begin{equation}
f_{0}\left(\xi\right)=\sum_{\epsilon_{\pmb{K}}>\epsilon_{F}}\dfrac{V^{2}}{\xi-\epsilon_{\pmb{K}}-f_{1}\left(\xi-\epsilon_{\pmb{K}}\right)}
\end{equation}

\begin{equation}
f_{1}\left(\xi\right)=\sum_{\sigma,\epsilon_{\pmb{k}}<\epsilon_{F}}\dfrac{2V^{2}}{\xi +\epsilon_{\pmb{k}}-V_{g}+\sigma B-f_{0}\left( \xi +\epsilon_{\pmb{k}}\right) }
\end{equation}

In the thermodynamic limit these equation can be written as,

\begin{equation}
f_{0}\left(\xi\right)=\int_{0}^{2t}\bigg \{ \rho(\omega) \dfrac{V^{2}}{\xi-\omega-f_{1}\left(\xi-\omega\right)}\bigg \} d\omega
\end{equation}

\begin{align}
f_{1}\left(\xi\right)&=\sum_{\sigma}\int_{-2t}^{0}\bigg \{ \rho(\omega)\times\nonumber\\
&\dfrac{2V^{2}}{\xi+\omega-V_{g}+\sigma B-f_{0}\left(\xi+\omega\right)}\bigg \}d\omega
\end{align}
where $\rho\left(\omega\right)$ is the density of states of the leads. It can be written as,
\begin{equation}
\rho\left(\omega\right)=\rho_{LC}(\omega)=\dfrac{1}{\pi\sqrt{4t^{2}-\omega^{2}}}
\label{eq:rho_LC}
\end{equation}
or
\begin{equation}
\rho(\omega)=\rho_{SC}(\omega)=\dfrac{\sqrt{4t^{2}-\omega^{2}}}{2\pi t^{2}}
\label{eq:rho_SC}
\end{equation}
that corresponds to a one dimensional linear chain, equation \eqref{eq:rho_LC}, or to two linear semi-chains, equation \eqref{eq:rho_SC}, depending on the geometry of the system.

\begin{figure}[h!]
\centering 
\includegraphics[width=\columnwidth]{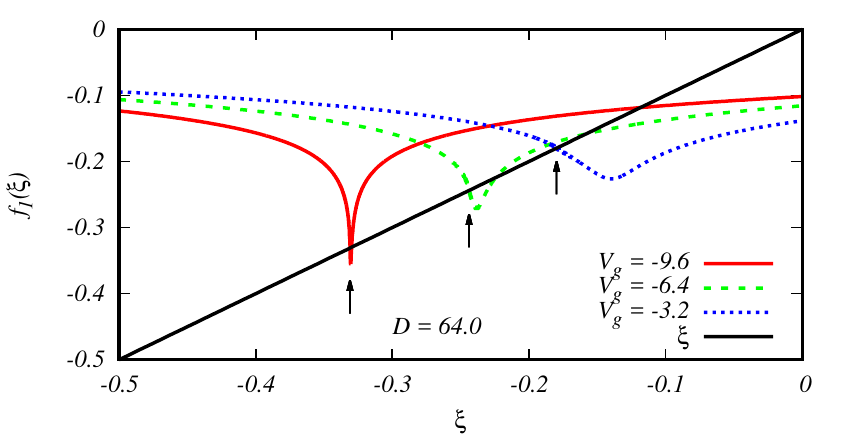}
	\caption{(Color online) The function $f_{1}\left(\xi\right)$ solved self-consistently for different 
	values of $V_g$ with a zero external magnetic field. The ground state energy solutions obtained by 
	ths POA approximation are given by intersections indicated by arrows.}
\label{figure10}
\end{figure}

The behavior of the function $f_{1}\left(\xi\right)$ is represented on Fig. \ref{figure10} for three values of $V_{g}$. The ground state solution corresponds to the lesser value of the intersection between the straight line and the $f_{1}\left(\xi\right)$ curves, that occurs on $\xi=\Delta E$. 
It can be shown that the derivative of the function $f_{1}\left(\xi\right)$ is singular at the point, $\Delta E=f_{1}\left(\Delta E\right)$, from which the energy is determined \cite{POA_1,POA_2}.
As we decrease $V_{g}$, the peak with a minimum value becomes sharper and other solutions with greater energy are possible. 
However we are interest only in the ground state energy of the system.

\bibliography{filtervf}

\begin{thebibliography}{75}
\bibitem [{not({\natexlab{b}})}]{note-variational}%
  \BibitemOpen
  \href@noop  \bibinfo {The expression for the
  variational Kondo temperature is given by $T_{K} = \{\left( D+2B \right) D
  \exp\left[\pi\left(V_{g}-B \right)/\left(2\Delta \right) \right] + B^2
  \}^\frac{1}{2}-B$, where $D=2t$ is the half-bandwidth. See
  Ref.~\onlinecite{Tosi2012}.}\BibitemShut {Stop}%
\end{thebibliography}

\end{document}